\documentclass[12pt]{article}
\usepackage{amssymb,epsfig,amsmath,amsfonts,graphicx}

\newcommand{\Tr}{\text{\,Tr\,}}
\newcommand{\eff}{\text{eff}}

\newcommand{\diag}{\text{\,diag\,}}
\newcommand{\LL}{\mathcal{L}}
\newcommand{\OO}{\mathcal{O}}
\newcommand{\PP}{\mathcal{P}}
\newcommand{\TeV}{\text{ TeV }}
\newcommand{\GeV}{\text{ GeV}}

\newcommand{\half}{\frac{1}{2}}
\newcommand{\hc}{\text{ h.c. }}
\newcommand{\identity}{{\rlap{1} \hskip 1.6pt \hbox{1}}}
\newcommand{\lsim}{\,\raise.3ex\hbox{$<$\kern-.75em\lower1ex\hbox{$\sim$}}\,}
\newcommand{\gsim}{\,\raise.3ex\hbox{$>$\kern-.75em\lower1ex\hbox{$\sim$}}\,}

\begin{document}

\begin{titlepage}
\begin{flushright}
HUTP-03/A022 \\
MIT-CTP-3377\\
hep-ph/0305275\\
\end{flushright}
\vskip 2cm
\begin{center}
{\large\bf What Precision Electroweak Physics Says\\ 
About the $SU(6)/Sp(6)$ Little Higgs}
\vskip 1cm
{\normalsize
Thomas Gregoire$^{(a)}$, David R. Smith$^{(b)}$ and Jay G. Wacker$^{(a)}$\\
\vskip 0.3cm

(a) Jefferson Physical Laboratory\\
Harvard University\\
Cambridge, MA 02138\\

\vskip 0.2in

(b) Center for Theoretical Physics\\
Massachusetts Institute of Technology\\
Cambridge, MA 02139\\
\vskip .1in}
\end{center}

\vskip .3cm

\begin{abstract}
We study precision electroweak constraints on the close 
cousin of the Littlest Higgs, the $SU(6)/Sp(6)$ model.
We identify a near-oblique limit in which the heavy $W'$ and $B'$
decouple from the light fermions, 
and then calculate oblique corrections, including 
one-loop contributions from
the extended top sector and the two Higgs doublets.
We find regions of parameter space that give acceptably small
precision electroweak corrections and only mild fine tuning
in the Higgs potential, 
and also find that the mass of the lightest Higgs boson is
relatively unconstrained by precision electroweak data.
The fermions from the extended top
sector can be as light as $\simeq 1$~TeV,
and the $W'$ can be as light as $\simeq 1.8$~TeV.  
We include an independent breaking scale for the $B'$,
which can still have a mass as low as a few hundred GeV. 

\end{abstract}

\end{titlepage}

%%%%
\section{Introduction}
In the Standard Model, the mass-squared parameter for the Higgs doublet 
is quadratically sensitive to the scale where new physics enters.   
While it is possible that there are cancellations
between the bare mass of the Higgs and its quantum corrections, or 
amongst the quantum corrections themselves, this
extreme sensitivity to cutoff scale physics makes these cancellations
increasingly delicate as the cutoff is raised. It is natural to
expect that the quantum corrections to the Higgs mass 
are not much larger than the vacuum expectation value of the Higgs, $v
= 247$ GeV, suggesting that 
new physics responsible for softening the radiative contributions
to the Higgs mass should appear near the TeV scale. 

Experiments performed at LEP and SLC are precise
up to one part in  $10^{-5}$, and provide 
indirect tests for physics beyond the Standard Model. 
These tests generally exclude models whose predictions 
deviate substantially from those of the Standard Model
electroweak sector.  Roughly speaking, corrections from new physics 
are restricted to be smaller than the 1-loop contributions 
of the Standard Model. Whenever new physics is 
proposed for stabilizing  the weak scale, it is important 
to evaluate its effects on precision electroweak observables.

The leading candidate for physics beyond the Standard Model is
the Minimal Supersymmetric Standard Model (MSSM).  
The supersymmetric partners of Standard Model fields
decouple quickly as their masses become larger than a few hundred GeV,
making their contributions to precision electroweak observables 
adequately small.   
The Standard Model quadratic divergences are cut-off at $M^2_{\text
  SUSY}$ the scale of the
superpartner masses, but have a logarithmic enhancement  
$\log \Lambda^2/M^2_{\text{SUSY}}$.  In many models these logarithmic
enhancements are $\OO(10^2)$, and the non-discovery of the
superpartners and of the lightest Higgs boson
requires some amount of fine tuning in the Higgs sector.  
Furthermore, for generic supersymmetry breaking
parameters, the superpartners mediate FCNCs at rates well beyond
current experimental bounds.  

Recently the little Higgs (LH) mechanism has emerged as a viable 
possibility for stabilizing the weak scale
\cite{Arkani-Hamed:2001nc,Arkani-Hamed:2002pa,Arkani-Hamed:2002qx,Gregoire:2002ra,Arkani-Hamed:2002qy,Low:2002ws,Kaplan:2003uc,Wacker:2002ar,Schmaltz:2002wx,Nelson:2003aj}.
In LH models the Higgs is a pseudo-Goldstone boson which is kept  
light by approximate non-linear symmetries.
At the weak scale, LH models contain 
the Standard Model with a weakly coupled Higgs and possibly several
other scalars.  At the TeV scale there are new states responsible
for canceling the quadratic divergences of the Standard Model. 
There are new vector bosons, scalars 
and fermions canceling the quadratic divergences from the gauge, 
scalars, and fermion interactions respectively. 
Little Higgs theories are described by
non-linear sigma models whose self-interactions become strongly
coupled at a scale $\Lambda \sim 10 - 15$ TeV where an ultraviolet 
completion is necessary to describe physics at higher scales.  

The low cutoff means that there is only a small
logarithmic enhancement to radiative corrections in the effective theory. Therefore
the new particles can be a factor of 5 -- 10 heavier than the
superpartners of the MSSM without making the fine tuning more severe
than in the MSSM.
This may explain why there has been no direct evidence
for the particles stabilizing the weak scale.
Furthermore there need not be any flavored particles at the TeV scale
in LH models, which can partially explain the absence of FCNCs.    Finally, in LH models
there is not in general an upper bound on the mass of the Higgs, so there is no
tension between the experimental lower bound on the Higgs mass and 
naturalness, as there is in the MSSM.

Precision electroweak constraints on LH models were first considered for
the $2 \times 2$ moose \cite{Arkani-Hamed:2001nc, Arkani-Hamed:2002pa} in \cite{Chivukula:2002ww} and for 
the ``Littlest Higgs'' \cite{Arkani-Hamed:2002qy} in \cite{Hewett:2002px,Csaki:2002qg}.
It was shown that some interactions in these
models can lead to significant contributions to precision electroweak
observables, but the constrained interactions are not the ones
responsible for stabilizing the weak scale.
In the model of \cite{Chang:2003un} an approximate custodial $SU(2)$
allowed for a certain limit where all the constraining physics decoupled
without making the states stabilizing the weak scale heavy.
We will examine this ``near-oblique limit'' in more
depth.

In this paper we show that the close cousin of the Littlest Higgs,
the $SU(6)/Sp(6)$ model \cite{Low:2002ws}, has regions of parameter space that
give small precision electroweak corrections even though
there is no obvious custodial $SU(2)$.  This is a two Higgs doublet
model in which the electroweak triplet scalar of the Littlest Higgs is
replaced by a neutral singlet.  An analysis of this model was recently
performed by \cite{Csaki:2003si}, but because it was done at 
tree-level, it neglected two significant 
contributions to precision electroweak observables -- loop 
contributions to the $S$ and $T$ parameters from the top and Higgs sectors.
We find that these contributions are large enough that it is important
to include them when constraining the model.  
%In this paper we will
%consider several different top sectors.  Their primary differences are 
%in the radiative corrections to the Higgs mass and also to 
%the oblique $S$ and $T$ parameters and these  effects make
%significant statements about whether a theory is constrained or not. 

We will consider a top sector whose radiative effects give the Higgs
a finite negative mass-squared.  Because this correction is
calculable in the effective theory,  naturalness considerations 
become quite straightforward.  The top sector's minimum contribution to the
Higgs mass-squared is 
\begin{eqnarray}
\label{Eq: Intro Top Rad}
\delta m^2 = \frac{81}{32 \pi^2} 
\left(\frac{m_{\text{top}}}{v \cos \beta}\right)^4 f^2
\simeq \frac{f^2}{4},
\end{eqnarray}
where we take $\tan \beta$, the ratio of
vacuum expectation values of the two Higgs doublets, to
be close to unity, as will be preferred by precision data.
Naturalness then motivates us to take $f \lsim 1$~TeV.
What is the lower bound on  $f$?  Cutoff suppressed higher dimension
operators give sizable contributions to precision electroweak
observables as the cutoff is lowered to $6~$TeV.
Here we adopt 0.7~TeV as a lower limit for $f$, corresponding
to a cutoff  $\lambda \sim 4\pi f \sim 9$~TeV. 

With such a low breaking scale, there is not a large separation between $v$
and $f$, and as one might expect, precision electroweak corrections
can be important.  Here we are interested in whether there are
regions of parameter space that satisfy the following criteria:
\begin{itemize}
\item the Higgs sector is natural 
\item there are only small contributions to precision electroweak observables 
\item all new particles are heavier than current experimental limits
\end{itemize} 
A full exploration of these regions 
would require a global fit to multiple parameters, a task we do not
undertake here. But we do find that parameters satisfying our criteria
exist.  More precisely, there are parameters for which 
the Higgs sector is only fine-tuned at the $10 -20$\%
level, while (i) oblique corrections to precision observables are
small and (ii) non-oblique corrections involving the light two
fermion generations are small.  Possible non-oblique corrections 
involving the third generation will be discussed only briefly, as
the correct interpretation of current precision data involving
the bottom quark is not completely clear.  Here 
we simply work under the assumption that 
an analysis in terms of $S$ and $T$ 
parameters is meaningful provided that the non-oblique
corrections associated with the light two generations 
are sufficiently small.

In Sec~\ref{Sec: Intro PEW} we discuss the near-oblique limit in which 
the above condition (ii) is
satisfied, and in Sec.~\ref{Sec: Overview} we summarize
the remaining corrections and outline the rest of the paper.

\subsection{The Near-Oblique Limit}
\label{Sec: Intro PEW}

Precision electroweak corrections fall into two categories,
non-oblique and oblique.  In the presence of non-oblique
corrections, extracting the  $S$ and $T$ parameters from experimental
results is complicated, which is why understanding precision electroweak 
constraints in LH models can become an intricate task.  Fortunately LH models 
that have a product gauge group structure have a simple limit where most
non-oblique corrections vanish
\cite{Hewett:2002px,Csaki:2002qg,Chang:2003un,Csaki:2003si,Kribs:2003yu}, 
so that the $S$ and $T$ parameters
can be interpreted in a clear manner.   It is not generally
possible to take a completely oblique limit, because in LH models the
physics of the third generation
tends to be different than that of the light generations, and for
this reason we call the limit ``near-oblique.''

In little Higgs models described by a product gauge group there is a
$W'$ and often a $B'$ that come from an enhanced electroweak
gauge sector. %$[SU(2)\times U(1)]^2$.  These heavy gauge bosons interact 
with the Standard Model fermions through interactions that are proportional to
the $SU(2)_L$ currents $j_F^{\mu\,a}$, and the $U(1)_Y$ current $j_F^\mu$, 
respectively.   The heavy gauge bosons also interact
with the Higgs fields, including the Goldstone bosons eaten by
the $W^\pm$ and $Z^0$, through the currents $j_H^{\mu\,a}$ and $j_H^\mu$, 
respectively.   Integrating out the $W'$ and the $B'$ generates 
$j_F j_F$, $j_F j_H$, and $j_H j_H$ current-current
interactions.   The $j_F j_F$ and $j_F j_H$ interactions give non-oblique
corrections that affect the extraction of $S$ and $T$ from various
precision electroweak results.  We now describe limits in which these 
non-oblique corrections vanish.

Consider first the four-Fermi interactions 
\begin{eqnarray}
\LL_\eff  \subset 
\frac{c_{FF}^L}{M^2} j^{\mu\,a}_F j^a_{F\,\mu}
+\frac{c_{FF}^Y}{M^2} j^{\mu}_F j_{F\,\mu}.
\end{eqnarray}
At low energies these modify  $G_F$ as well as other observables.  The 
coefficient of the $W'$ coupling is related to the ratio of the two $SU(2)$ 
gauge couplings $g_1$ and $g_2$.  If the Standard Model
fermions are charged under $SU(2)_1$, then the coupling of
the $W'$ to the fermions is proportional to $g_1/g_2$.  As $g_2$
becomes larger than $g_1$ the $W'$ decouples from the Standard Model 
fermions and $c_{FF}^L$ eventually becomes negligible.   
The ratio of couplings does not need to be 
extreme for $c_{FF}^L$ to be small enough for our purposes, 
with $g_2 \sim 2 - 3$ sufficiently large. 

The $U(1)_Y$ current-current interaction comes from integrating
out the $B'$.  The interactions of the $B'$ are more model-dependent 
than those of the $W'$ because of freedom in assigning the fermion
charges under the each $U(1)$.     Furthermore, the $U(1)_Y$ quadratic
divergence is only marginally relevant to the naturalness of the Higgs 
potential in little Higgs theories: it does not become significant until 
$\Lambda \sim 10 \TeV$, which is typically where new physics is expected 
to be present.  This leads us to be open to the possibility that only 
$U(1)_Y$ is gauged, rather than a product of $U(1)$'s.  If a product of 
$U(1)$'s is gauged, charging the light Standard Model fermions equally under 
both factors yields couplings to the $B'$ that vanish as the two $U(1)$ 
couplings $g'_1$ and $g'_2$ become equal.

The $j_H j_F$ interactions that contribute to precision electroweak 
constraints are
\begin{equation}
\LL_\eff \subset  
\frac{1}{M^2} j_H^\mu j_{F\,\mu}
+\frac{1}{M^2} j_H^{\mu\,a} j_{F\,\mu}^a
=\frac{1}{M^2} j_\omega^\mu j_{F\,\mu}
+\frac{1}{M^2} j_\omega^{\mu\,a} j_{F\,\mu}^a +\cdots
\end{equation}
where $j^\mu_H = i h^{\dagger} \overleftrightarrow{D}^{\mu} h$,
$j^{a\,\mu}_H = i h^{\dagger} \sigma^a\overleftrightarrow{D}^{\mu} h$ 
are the Higgs currents,
and $j^{a\,\mu}_\omega = \frac{i}{2} v^2 
\Tr \sigma^3 \omega^\dagger\overleftrightarrow{D}^{\mu}\omega$
and $j^\mu_\omega = \frac{i}{2} v^2 
\Tr \sigma^a \omega \overleftrightarrow{D}^{\mu}\omega^\dagger$
are the currents involving only the Goldstone bosons.
In unitary gauge these currents simply become
$j^\mu_\omega \simeq v^2 Z^0{}^\mu$ and $j^{a\,\mu}_\omega 
\simeq v^2 (W^\pm{}^\mu, Z^0{}^\mu)$.
These modify the $Z^0$ and $W^\pm$ couplings to the 
Standard Model fermions and affect $Z^0$-pole observables
as well as low energy tests.
However, in the limits just described, the $W'$ and $B'$
decouple from the Standard Model fermion fields, and 
the above operators will not be generated with large coefficients.
Moreover, the $B'$ couples as $({g_1'}^2 - {g_2'}^2)$ to the Higgs current,
so the same limit that decouples the $B'$ from fermions decouples it
from the Higgs current as well.

The $j_H j_H$ operators give oblique corrections because the 
constrained interactions  
include just the Goldstone modes.   
We should check that in taking the near-oblique limit 
we have not made these oblique corrections large. 
The $(j_H^a)^2$ interaction is independent
of the Goldstone bosons and only renormalizes $v^2$
by a finite amount and thus has no limits placed on it.
On the other hand, the current-current interaction mediated by the $B'$, 
\begin{equation}
\LL_\eff \subset \frac{c_{T\;B'}}{M^2} |j_H^\mu |^2 =
\frac{c_{T\;B'}}{M^2} |j_{\omega}^\mu|^2 +\cdots 
\label{eq:bint}
\end{equation}
has observable effects.
In unitary gauge this is just an extra mass term for the $Z^0$.
Fortunately, in the near-oblique limit described above, the
Higgs current decouples from the $B'$ and this oblique correction
becomes small simultaneously.

In summary, the near-oblique limit fixes $\tan \theta \equiv g_1/g_2$ to
be somewhat small and $\tan \theta' \equiv g'_1/g'_2$ to
be roughly unity.   This limit also  ensures that the
$W'$ and $B'$ do not mediate large oblique corrections.

\subsection{Overview and Summary}
\label{Sec: Overview}

\begin{table}
\renewcommand{\arraystretch}{1.3}
\begin{center}
\begin{tabular}{|c|l|c|c|}
\hline
Parameter& Relevance & Ref. Val. 1& Ref. Val. 2 \\
\hline
\hline
$f$& Breaking scale of nl$\sigma$m& 700 GeV& 700 GeV\\
$F$& Breakings scale for $B'$ & 2 TeV& 2 TeV\\
\hline
$\tan\theta$& $SU(2)$ mixing angle & $\frac{1}{5}$&$\frac{1}{5}$\\
$\tan \theta'$& $U(1)$ mixing angle & $\frac{9}{10}$&$\frac{9}{10}$\\
\hline
$\cos^2\vartheta_U$& Singlet top mixing angle& $\frac{2}{3}$&$\frac{3}{4}$\\
$\cos^2\vartheta_Q$& Doublet top mixing angle& $\frac{2}{3}$&$\frac{3}{5}$ \\
\hline
$\lambda$& Higgs quartic coupling& 3.0 & 0.5\\
$\Delta m^2$& Gauge and scalar contribution to Higgs mass
& (550 GeV)${}^2$& (550 GeV)${}^2$\\
\hline
\end{tabular}\\
\end{center}
\caption{
\label{Table: Parameters}
Important parameters of $SU(6)/Sp(6)$.
}
\end{table}
What are the parameters that determine the physics of the
$SU(6)/Sp(6)$ model? We list a complete set in Table~\ref{Table: Parameters}.  
As will be discussed in Sec.~\ref{sec:gauge}, the $U(1)^2$
gauge symmetry is contained in $U(6)$ rather than $SU(6)$, so
along with the $SU(6)$ breaking scale $f$, there is an additional
scale $F$ associated with the breaking $U(6)\rightarrow SU(6)$.
In the gauge sector, there are two mixing angles $\theta$ and
$\theta'$ already introduced.  In the top sector that will be
considered in Sec.~\ref{Sec: Top}, there is a second pair of mixing angles
$\vartheta_Q$ and $\vartheta_U$, which describe the mixing of the
third generation quarks with additional vector-like quarks.  Finally,
there are two additional parameters in the Higgs sector, which may be
taken to be the quartic coupling $\lambda$ and $\Delta m^2$, which is the radiative
correction to the Higgs masses, leaving out the top contribution. The
parameter $\Delta m^2$
has cutoff dependence and so cannot be calculated in the effective theory,
but the top contribution, $\delta m^2$, is finite 
and calculable in terms of the parameters of the theory. 

For illustrative purposes, in Table~\ref{Table: Parameters} we also give
two sample sets of parameter values that give
small precision electroweak corrections.  The breaking scale $f$ is
taken to be as low as we are comfortable taking it given its relation
to the cutoff.  The only consequence of $F$ is to give additional mass to the
$B'$, thereby relaxing constraints associated with it.  The mixing
angles in the gauge sector are chosen to be close to the near-oblique
limit; $\tan \theta$ cannot be taken arbitrarily small or else $g_2$
becomes non-perturbative, but the value we have taken is sufficiently
small to adequately suppress the non-oblique corrections.  
Both sets of values chosen for the mixing
angles of the top sector essentially minimize the radiative correction 
$\delta m^2$, as is favorable for naturalness (for both sets of parameters 
we get $\delta m^2=(300~\rm{GeV})^2$).  
These choices also produce relatively small oblique corrections coming 
from the top sector.  

The first set gives a positive contribution to $T$ from the top sector.
In this case a larger value for the quartic coupling $\lambda$ is 
somewhat preferred because the Higgs contribution to $T$ is negative
and grows in magnitude with $\lambda$.  These larger values of
$\lambda$ also reduce the fine tuning in the Higgs sector.  On the
other hand for the second set, the contribution to $T$ from the top
sector is negative, and in this case smaller values for $\lambda$ 
(and for the mass of the lightest Higgs boson) are equally consistent
with precision data. 

Finally, $\Delta m^2$ is taken to be roughly the size expected from the 
one-loop logarithmically divergent contributions to the Higgs masses.  By
taking it to be somewhat larger than $\delta m^2$, we get
 $\tan \beta \simeq 1$ for the ratio of Higgs vevs, 
avoiding large custodial $SU(2)$ violation in the non-linear sigma model 
self-interactions.  On the other hand, since it is  
only moderately large the Higgs sector will be not severely fine-tuned.

As will become clearer in Sec.~\ref{Sec: EWSB}, a reasonable measure of the 
fine tuning in the Higgs sector  is
\begin{eqnarray}
\text{fine tuning } \simeq\; \kappa \equiv  
\frac{\lambda v^2}{4 \Delta m^2 -2 \delta m^2}
\label{Eq: Intro FT}
\end{eqnarray}
and the reference values given in Table~\ref{Table: Parameters}
give $\kappa \simeq 0.18$ and 0.03, respectively, corresponding to a 
Higgs sector tuned at about the 20\% and  3\% levels.  
Meanwhile, the contributions to the $S$ and $T$ parameters for the reference 
values are reasonably small: $(0.13,0.13)$ and $(0.08, 0.13)$. 

The outline of the rest of the paper is as follows.
In Section 2 we review the $SU(6)/Sp(6)$ little Higgs model and discuss
our conventions, which differ from those of \cite{Low:2002ws,Csaki:2003si} in 
several ways. We also consider a few important modifications of the
model as presented in \cite{Low:2002ws}.  For instance, we allow for
the additional breaking scale $F$ that makes the $B'$ heavier, and
consider a top sector
that removes both one and two loop quadratic divergences to
the Higgs mass,
leaving the correction from the top sector calculable.   This
``full six-plet'' top sector may be easier to embed into a
composite Higgs model along the lines of \cite{Nelson:2003aj}.
In Section 2 we also study the Higgs sector in detail.
The mass terms  $m_1^2 |h_1|^2 + m_2^2 |h_2|^2$
are generated by radiative corrections within the effective theory,  
and the structure of radiative corrections tends to give 
$\tan \beta = m_1/m_2 \sim 1$.

In Sec.~\ref{Sec: PEW} we consider precision electroweak constraints
on the model.  In Sec.~\ref{Sec: PEWCurrents} we discuss non-oblique
corrections and apply the near-oblique limit discussed in 
Sec.~\ref{Sec: Intro PEW}.   In Sec.~\ref{Sec: Oblique} we calculate oblique 
corrections coming from four sources:  integrating out the heavy gauge bosons, 
the non-linear sigma model self-interactions, loop corrections from the
Higgs sector, and loop corrections from the top sector.
We find sizable negative contributions to $T$ and positive contributions to 
$S$ from the Higgs sector, which are both helpful in light of the positive 
contributions to $T$ from the other sources. In Sec.~\ref{Sec: 3rd Gen} 
non-oblique corrections involving the third generation fermions are briefly
discussed, and in  Sec.~\ref{Sec: B' Production} issues involving $B'$
production are mentioned.  In Sec.~\ref{Sec: Conc} we analyze our
results, focusing on implications for the Higgs sector, and give our 
conclusions.
%%%%

%%%%
\section{The $SU(6)/Sp(6)$ Little Higgs}
\setcounter{equation}{0}
\renewcommand{\theequation}{\thesection.\arabic{equation}}

In this section we review the $SU(6)/Sp(6)$ little Higgs model.
We also investigate possible modifications to the model as presented
in \cite{Low:2002ws}, and aspects of the model that were 
either not discussed in \cite{Low:2002ws} or not explored thoroughly.    
For instance, we consider the possibility that the $B'$ mass is
independent of the $SU(6)$ breaking scale, and also note the presence of an
axion in the theory, and give an 
example of an operator that can generate a mass for the axion.  
We identify global symmetries preserved in the vacuum 
that are helpful in thinking about the
structure of radiative corrections to the Higgs potential, and study
the Higgs spectrum in more detail.
Unlike \cite{Low:2002ws}, we allow the light two generations of
Standard Model fermions to be charged under both $U(1)$ gauge symmetries, 
and we consider some of the implications of such a setup.  Finally,
we consider a new setup for the third generation that gives
finite radiative corrections to the Higgs mass-squared.

\subsection{Basic Structure}
\label{sec:gauge}
The $SU(6)/Sp(6)$ little Higgs is a gauged non-linear sigma model 
with $\Sigma = - \Sigma^T$ and $\Sigma \Sigma^\dagger = \identity$.
$\Sigma$ transforms under global $SU(6)$ transformations $V$ as
\begin{eqnarray}
\Sigma \rightarrow V \Sigma V^T.
\end{eqnarray}
We choose a basis where the vacuum is
\begin{eqnarray}
\label{Eq: Vacuum}
\langle \Sigma \rangle = \Sigma_0 = 
\left(
\begin{array}{cccc} 0 & \identity_2&0&0\\ - \identity_2&0&0 & 0 \\0&0&0&1
\\0&0&-1&0\end{array}
\right) .
\end{eqnarray}
This basis is different than that chosen in \cite{Low:2002ws}; we
choose it because it more clearly exhibits
two separate $SU(2)$ symmetries that are preserved in the vacuum and 
that are important for constraining radiative corrections.

The generators of $SU(6)$ can be separated into broken and
unbroken ones,
\begin{eqnarray}
[T_x, \Sigma_0] =0
\hspace{0.5in}
\{X_m, \Sigma_0\}  =0,
\end{eqnarray}
where $T_x$ form an $Sp(6)$ algebra and $X_m$ are
the broken generators in $SU(6)/Sp(6)$.
The linearized fluctuations around the vacuum, $\pi$  appear
in 
\begin{eqnarray}
\Sigma = 
\exp\Big(i \frac{\pi}{f}\Big) \Sigma_0 \exp\Big(i \frac{\pi^T}{f}\Big)
= \exp\Big(2 i \frac{\pi}{f}\Big) \Sigma_0
\end{eqnarray}
where $\pi = \pi^m X_m$.  

An $SU(2)^2$ subgroup of $SU(6)$ is gauged, with generators
\begin{eqnarray}
T_1^a =
\half
\left( \begin{array}{ccc} \sigma^a&&\\ & 0_2 &\\
&&0_2
\end{array}
\right)
\hspace{0.5in}
T_2^a =
- \half
\left( \begin{array}{ccc} 0_2&&\\ & \sigma^a{}^* &\\
&&0_2 
\end{array}
\right).
\end{eqnarray}
The vacuum  breaks the gauge sector down to the diagonal 
$SU(2)$, which we identify as $SU(2)_L$ of the Standard Model.  
The physics of hypercharge is more subtle in little Higgs models
because the $U(1)_Y$ quadratic divergence to the Higgs mass does
not spoil naturalness until scales $\Lambda \gsim 10 \TeV$.  
Hence a reasonable approach is simply to gauge hypercharge alone
and live with the relatively small quadratic divergence,
possibly allowing for easier embedding into ultraviolet completions.
In this scenario the hypercharge generator is
\begin{eqnarray}
Y =
\half
\left(
\begin{array}{cccc}
0_2&&&\\ & 0_2 &&\\ &&1&\\ &&&-1 \end{array} \right).
\end{eqnarray}
If one insists on canceling the one loop quadratic divergence 
associated with hypercharge,
the easiest way is by gauging $U(1)^2$ with  generators
\begin{eqnarray}
Y_1 =
- \half
\left( 
\begin{array}{cccc}
0_2&&&\\
& 0_2 &&\\
&&1&\\
&&&0
\end{array}
\right)
\hspace{0.5in}
Y_2 =
 \half 
\left(
\begin{array}{cccc}
0_2&&&\\
& 0_2 &&\\
&&0&\\
&&&1
\end{array}
\right) .
\end{eqnarray}
The vacuum breaks $U(1)^2$ down to the diagonal $U(1)_Y$.
Because the $U(1)^2$ lives inside $U(6) = U(1)_0\times SU(6)$ 
rather than $SU(6)$ alone
we have implicitly introduced another Goldstone boson and
an additional breaking scale associated with $U(1)_0$.
The Standard Model quadratic divergence from hypercharge
is cutoff at the mass of the $B'$.  In our analysis, we will
explore the consequences of gauging a product of $U(1)$'s, but
one should also keep in mind the simpler (and less constrained) possibility
of only gauging $U(1)_Y$. 

The vacuum respects  $SU(2)_R$, $SU(2)_H$ and $U(1)_{PQ}$ global
symmetries that are
approximate symmetries of the full theory.  These have generators  
\begin{eqnarray}
T_R^a = \left(
\begin{array}{ccc}0_2\\
&0_2\\
&&\sigma^a
\end{array}\right)
\hspace{0.3in}
T_H^a = \left(
\begin{array}{cc}
\sigma^a\otimes \identity_2\\
&0_2
\end{array}\right)
\hspace{0.3in}
Q_{PQ}= T^3_H = \left(
\begin{array}{ccc}
\identity_2\\
& - \identity_2\\
&&0_2
\end{array}\right)
\end{eqnarray}
The $SU(2)$ gauge  generators commute with $SU(2)_R$ but not with $SU(2)_H$
while the $U(1)$ gauge generators do not commute with $SU(2)_R$ but 
commute with $SU(2)_H$.

The massive vector bosons $W'$ and $B'$ eat $\mathbf{3_0} +\mathbf{1_0}$ 
Goldstone bosons leaving 10 physical pseudo-Goldstone bosons:
$\eta \sim \mathbf{1_0}$, $h_1 \sim \mathbf{2_{+\half}}$,
and $h_2\sim \mathbf{2_{+\half}}$, and finally $a \sim \mathbf{1_0}$, 
which is an axion.  In unitary gauge the modes reside within $\pi$ as
\begin{eqnarray}
\pi = \frac{1}{\sqrt{2}} \left( \begin{array}{cccc}
\frac{1}{\sqrt{2}} a \identity_2&\epsilon_2 \eta & h_1 & h_2^*\\
-\epsilon_2 \eta^* & \frac{1}{\sqrt{2}}a \identity_2 &- h_2 & h_1^*\\
h_1^\dagger& - h_2^\dagger &0&0\\
h_2^T& h_1^T &0&0
\end{array}
\right),
\end{eqnarray}
where $\epsilon_2$ is the $2\times 2$ antisymmetric tensor.
Note that both $SU(2)_H$ and $SU(2)_R$ rotate one Higgs doublet into
the other.  

The kinetic terms for the non-linear sigma model field are
\begin{eqnarray}
\label{Eq: NLSM Kin}
\LL_{\text{nl$\sigma$m Kin}} 
= \frac{f^2}{8} \Tr  D_\mu \Sigma D^\mu \Sigma^\dagger
+  \frac{F^2}{2} |D_\mu \det \Sigma|^2
+ \cdots,
\end{eqnarray}
where $\cdots$ represents higher order operators in the
Lagrangian, and where the covariant derivative is given as
\begin{eqnarray}
\nonumber
D_\mu \Sigma &=& \partial_\mu \Sigma 
-i g_1 W^a_{1\,\mu} ( T_1^a \Sigma + \Sigma T^{a\,T}_1)
-i g_2 W^a_{2\,\mu} ( T_2^a \Sigma + \Sigma T^{a\,T}_2)\\
&&\hspace{0.35in}
-i g'_1 B_{1\,\mu} ( Y_1 \Sigma + \Sigma Y_1)
-i g'_2 B_{2\,\mu} ( Y_2 \Sigma + \Sigma Y_2),\\ 
D_\mu \det \Sigma &=& \partial _\mu \det \Sigma  
- g'_1 B_{1\,\mu}  \det \Sigma + g'_2 B_{2\,\mu}\det \Sigma.
\end{eqnarray}
The second term is Eq. \ref{Eq: NLSM Kin} is another 
$SU(6)$ invariant kinetic term that can be interpreted
as an additional breaking of the $U(1)\subset U(6)$. 
There is no \begin{it}a priori\end{it} size associated
with it so it can in principle be large.
The higher order terms in Eq. \ref{Eq: NLSM Kin} are of
the form
\begin{eqnarray}
\label{Eq: NLSM Kin 4D}
\frac{1}{16 \pi^2} \Tr| D_{[\mu} D_{\nu ]} \Sigma |^2 + \cdots
\end{eqnarray}
These operators typically contribute acceptably small
oblique corrections.

\subsection{Gauge Boson Masses and Couplings}

At lowest order in the linearized fluctuations of the non-linear sigma 
model field the kinetic term contains masses for the vector bosons.  
The Standard Model $SU(2)_L\times U(1)_Y$ gauge couplings are
\begin{eqnarray}
g^{-2} = g_1^{-2} + g_2^{-2} 
\hspace{0.5in}
g'{}^{-2} = g'_1{}^{-2} + g'_2{}^{-2}. 
\end{eqnarray}
The gauge bosons can be diagonalized with the transformations
\begin{eqnarray}
\nonumber
W^a &=& \cos\theta W_1^a - \sin\theta W_2^a
\hspace{0.5in}
W'{}^a= \sin \theta W_1^a + \cos \theta W_2^a\\
\nonumber
B &=& \cos\theta' B_1 - \sin\theta' B_2 
\hspace{0.5in} 
B'= \sin \theta' B_1 + \cos \theta'  B_2,
\end{eqnarray}
where the mixing angles are related to the couplings by
\begin{eqnarray}
\nonumber
\cos \theta &=& g/g_1 \hspace{0.5in} \sin \theta = g/g_2 \\
\cos \theta' &=& g'/g'_1 \hspace{0.5in} \sin \theta' = g'/g'_2. 
\end{eqnarray}

The masses for the vectors can be written in terms of the electroweak gauge
couplings and mixing angles:
\begin{eqnarray*}
m^2{}_{W'}  =  \frac{ g^2 f^2}{\sin^2 2\theta} 
\hspace{0.5in}
m^2{}_{B'} =  \frac{ g'{}^2 \bar{f}^2}{2\sin^2 2 \theta'},
\end{eqnarray*}
where $\bar{f}^2 = f^2 + F^2$.
The near oblique limit discussed earlier has $\theta' \simeq
\frac{\pi}{4}$
and $\theta \ll 1$.  
The $B'$, the mode that cancels the quadratic divergence of 
the $B$, receives additional mass from the second breaking
scale, $F$, but is still rather light in the near-oblique limit:
\begin{eqnarray}
\label{Eq: B' Mass}
m_{B'} \simeq 375 \GeV\; \left(\frac{\bar{f}}{2 \TeV}\right).
\end{eqnarray}
In order for such a light $B'$ to have 
%the absence of the extra $U(1)$ breaking ($F\rightarrow 0$), $B'$
%is so light that in order for the $B'$ to have 
evaded direct  Drell-Yan production searches 
the $B'$ must couple only weakly
to the light Standard Model fermions.  In Sec. \ref{Sec: Fermions} we
discuss the couplings of the Standard Model fields to the $B'$, and
we briefly consider the question of production constraints 
in Sec. \ref{Sec: B' Production}. 
Although the $W'$ mass increases as the near oblique limit is
approached, it is still likely to be moderately light because $f$ is so small, 
\begin{eqnarray}
m_{W'} \simeq 1.8 \TeV 
\left( \frac{f}{700 \GeV}\right)
\left( \frac{\csc \theta}{5}\right).
\label{eq:wprimemass}
\end{eqnarray}

The Higgs boson couples to these gauge bosons through the currents
\begin{eqnarray}
\nonumber
j^{\mu\;a}_{W'} & = &  
g \cot{2\theta}
(
i h_1^\dag \sigma^a \overleftrightarrow{D}^\mu h_1
+ i h_2^\dag \sigma^a \overleftrightarrow{D}^\mu h_2)
= g \cot 2\theta j_{\omega}^{\mu\;a} + \cdots
\\
j^{\mu}_{B'} & = & 
g' \cot{2\theta'}
(i h_1^\dagger \overleftrightarrow{D}^\mu h_1
+i h_2^\dagger \overleftrightarrow{D}^\mu h_2)
=g' \cot 2 \theta' j_{\omega}^{\mu} + \cdots
\label{Eq: Current Interactions}
\end{eqnarray}
where $D_\mu$ is the Standard Model covariant derivative
and $j_{\omega}^{\mu\;a}$ and $j_{\omega}^{\mu}$ are the
the $SU(2)_L$ and $U(1)_Y$ currents for the eaten Goldstone bosons, 
respectively.  
  
\subsubsection*{Radiative Corrections}

Radiative corrections in little Higgs models are most readily computed
with the Coleman-Weinberg potential, where one turns on a background
value for $\Sigma$.
The one loop quadratically divergent contribution to the scalar potential 
from the gauge sector is
\begin{eqnarray}
V_{\eff} = \frac{3}{32 \pi^2} \Lambda ^2 \Tr M^2[\Sigma]
\end{eqnarray}
where $M^2[\Sigma]$ is the mass matrix of the gauge bosons in the background
of the little Higgs.
The $SU(2)$ gauge sector gives a quadratic divergence of the form
\begin{eqnarray}
V_{\eff}= 
c_1 g_1^2 f^4 \Tr \PP_1 \Sigma \PP_1 \Sigma^\dagger
+c_2 g_2^2 f^4 \Tr \PP_2 \Sigma \PP_2 \Sigma^\dagger,
\end{eqnarray}
where $\PP_1 = \diag( \identity_2,0_2,0_2)$ and 
$\PP_2 = \diag( 0_2, \identity_2,0_2)$ are matrices arising from
the sum over the $SU(2)$ generators and $c_{1,2}$ are unknown
$\OO(1)$ coefficients that depend on the details of how
the ultraviolet physics cuts off the gauge quadratic divergences.  
%quadratic divergence, $\Lambda^2 \simeq c_{1,2}(4 \pi f)^2$.  
In terms of the linearized modes we have
\begin{eqnarray}
\label{Eq: Gauge Potential}
V_{\eff}=
c_1 g_1^2 f^4 \Tr \PP_1 e^{2i \pi/f} \PP_2 e^{-2i \pi/f}
+c_2 g_2^2 f^4 \Tr \PP_2 e^{2i \pi/f} \PP_1 e^{-2i \pi/f} .
\end{eqnarray}
Although it is not immediately apparent, these interactions
keep both the of the Higgs doublets light while giving the 
singlet a TeV-scale mass.  The naive sign for $c_1$ and $c_2$, based on the 
the gauge boson loop contributions, gives a local maximum of the
potential around $\Sigma_0$. However, since the sign is 
cut-off dependent, we will simply assume that $\Sigma_0$ gives 
a local minimum instead.
%note that if it had been a saddle point,
%such a statement would not be possible to make.  
We will explore the physics of
this potential in the next section.  Meanwhile, the $U(1)$ quadratic 
divergences in the case when there are two 
$U(1)$'s gauged vanish:
\begin{eqnarray}
V_{\eff}=
c'_1 g_1'{}^2 f^4 \Tr Y_1 e^{2i \pi/f} Y_2 e^{-2i \pi/f}
+c'_2 g_2'{}^2 f^4 \Tr Y_2 e^{2i \pi/f} Y_1 e^{-2i \pi/f} = 0.
\end{eqnarray}

The full one loop Coleman-Weinberg potential gives logarithmically
divergent and finite contributions to the little Higgs masses that are
appropriately small from the point of view of naturalness.  
The potential  generated for the Higgs doublets is
\begin{eqnarray}
V_{\eff} = \Big(\frac{3 g^2 M^2_{W'}}{8 \pi^2} 
\log \frac{\Lambda^2}{M^2_{W'}}  
+ \frac{3 g'{}^2 M^2_{B'}}{8 \pi^2}
\log \frac{\Lambda^2}{M^2_{B'}}\Big) (|h_1|^2 +|h_2|^2)
+ \OO(|h|^4) .
\label{Eq: gaugerad}
\end{eqnarray}
The $SU(2)_R$ and $SU(2)_H$ symmetries guarantee that the Higgs doublets 
have the same radiatively generated mass.

There are also two-loop  quadratic divergences which are parametrically
the same size as the one loop pieces, but
are not reliably calculated inside the effective theory, since the
majority of the contribution is coming from cut-off scale physics.  
Hence the physics can be encoded in  soft masses-squared that 
are of order $v^2$.  Provided the $SU(2)_{H,R}$ symmetries
are preserved in the ultraviolet theory, these incalculable contributions
will also be $SU(2)_{H,R}$ symmetric. 

\subsection{Scalar Masses and Interactions}
\label{sec:sc}
When expanded to leading in order in the scalars, 
the potential of Eq.~\ref{Eq: Gauge Potential} is
\begin{eqnarray}
\label{Eq: Quartic1}
 V= \frac{\lambda_1}{2} |  f \eta + \frac{i}{2} h_1^\dagger h_2|^2
 + \frac{\lambda_2}{2} |  f \eta - \frac{i}{2} h_1^\dagger h_2|^2 + \cdots ,
\end{eqnarray}
where we have replaced $c_i g_i^2$ with $\lambda_i$ since these
are incalculable in the low energy theory.

After integrating out $\eta$ the low energy potential for
the Higgs is
\begin{eqnarray}
\label{Eq: Quartic2}
V_{\text{Quartic}} = \lambda |h_1^\dagger h_2|^2,
\end{eqnarray}
with the quartic coupling given related to the high energy couplings through
\begin{eqnarray}
\lambda^{-1} = \lambda_1^{-1} + \lambda_2^{-1}
\hspace{0.6in} 
\tan^2 \vartheta_\lambda = \frac{\lambda_1}{\lambda_2}.
\end{eqnarray}
Taking $c_i =1$, the quartic coupling is given by
\begin{eqnarray}
\lambda = g^2 
\hspace{0.6in}
\tan \vartheta_\lambda  = \tan \theta,
\end{eqnarray}
but since the actual values are highly sensitive to cutoff scale
physics,
all that can be inferred is that it is unnatural to have 
$\lambda \ll g^2$.  The mass of the singlet $\eta$ is
\begin{eqnarray} 
m^2_\eta =  \frac{2 \lambda f^2}{\sin^2 2 \vartheta_\lambda} .
\end{eqnarray}

The light axion $a$, does not pick up a mass
through any mechanism that has been discussed so far.
It is a Goldstone boson of the broken symmetry
\begin{eqnarray}
Q_A = 
\left(
\begin{array}{ccc} \identity_2\\&\identity_2\\ &&0\end{array}
\right). 
\end{eqnarray}
The axion can be quite light without having significant phenomenological 
implications.   There are several ways of giving it a mass; 
one possible operator is
\begin{eqnarray}
\OO_{\text{Axion Mass}} = c_A f^4 
\Sigma_{\alpha \alpha'}
\Sigma_{\beta \beta'}\epsilon^{\alpha \beta} \epsilon^{\alpha' \beta'}
+\hc \simeq  
v^4 \sin^2(2 \beta ) \cos a/f +\cdots,
\end{eqnarray}
where $\alpha,\beta$ and $\alpha',\beta'$ run over $SU(2)_1$ indices.
This term does not induce one loop quadratic divergences in  the Higgs masses,
but gives the axion a mass after electroweak symmetry breaking of 
$\OO(v^4/f^2)$.  

\subsubsection*{Radiative Corrections}

Because the quartic potential in Eq.~(\ref{Eq: Quartic1}) has $\OO(1)$ coefficients,
one might worry that it destabilizes the weak scale, but 
if  either $\lambda_1$ or $\lambda_2$ vanishes then there is a global $SU(4)$ 
acting on the fields as
\begin{eqnarray}
\delta h_1 \approx \epsilon_1 
\hspace{0.4in} 
\delta h_2 \approx \epsilon_2 
\hspace{0.4in} 
\delta \eta \approx -\frac{i}{2 f} ( \epsilon_1^\dagger h_2 + h_1^\dagger \epsilon_2)\\
\delta h_1 \approx \epsilon'_1 
\hspace{0.4in} 
\delta h_2 \approx \epsilon'_2 
\hspace{0.4in} 
\delta \eta \approx \frac{i}{2 f} ( \epsilon'_1{}^\dagger h_2 + h_1^\dagger \epsilon'_2) .
\end{eqnarray}
Either one of these symmetries is sufficient to keep both Higgs doublets
as exact Goldstones.  Only through the combination of $\lambda_1$ and
$\lambda_2$ is there sufficient symmetry breaking to generate a mass
for the Higgs doublets, so a quadratic divergence arises only at two loops. 
%quadratic divergence that is parametrically of the same size as
%the gauge contributions to the little Higgs masses.   
The one-loop contributions to the Higgs masses are of the form 
\begin{eqnarray}
V_\eff = \frac{ \lambda m_\eta^2}{8 \pi^2} 
 \log \frac{\Lambda^2}{m^2_\eta} \big(|h_1|^2 + |h_2|^2\big) .
\label{Eq: scalarrad}
\end{eqnarray}
Notice that the $SU(2)_H$ symmetry remains unbroken and the low energy
quadratic divergence is cutoff at $m_\eta^2$, as expected.  
There are two-loop quadratically divergent contributions of comparable size.

The maximum size of the quartic coupling we are willing to consider is 
roughly determined by naturalness, although another concern is that $\lambda$ 
should not hit its Landau pole below the cutoff.   Focusing on  naturalness,
a reasonable limit is given by requiring 
$\Delta m^2_{\text{Scalar}}\lsim f^2$,
\begin{eqnarray}
\Delta m^2_{\text{Max Scalar}}
\simeq \frac{ 2 \lambda^2 f^2}{8 \pi^2}
 \log \frac{(4\pi)^2}{2 \lambda} \lsim f^2.
\end{eqnarray}
This leads to an approximate bound $\lambda \lsim 4$.

\subsection{First and Second Generation Fermions}
\label{Sec: Fermions}

Since the first and second generation fermions couple extremely weakly 
to the Higgs sector, we can simply write down standard Yukawa
couplings to the linearized Higgs doublets with out destabilizing
the weak scale. 
%The Yukawa couplings of the light two generations are treated as
%the only source of flavor in the low energy theory which ensures
%that the light fermions only couple to one combination of Higgs.
To avoid excessive FCNCs in the low energy theory we imagine that 
the light fermions only couple to a single linear combination of 
the two Higgs doublets. 
There are many ways to covariantize the Yukawa couplings, depending on the 
charge assignments of the light fermions.  These different approaches differ
only by irrelevant operators and so the differences are not important
for most collider phenomenology,  but might have implications for flavor 
physics.  

The Standard Model fermion doublets are charged only under $SU(2)_1$.  
Their coupling to the heavy gauge bosons is
\begin{eqnarray}
\LL_{\text{Int}} = g \tan \theta \; W'{}^a_\mu \, j_{\text{F}}^\mu{}_a.
\end{eqnarray}
In the near oblique limit we have $g_2 \rightarrow \infty$ and 
$\theta \rightarrow 0$, and the TeV scale gauge bosons decouple from the 
Standard Model fermions.  Notice that gauge invariance does not determine 
the couplings of the Standard Model fermions to the $W'$ because it is 
associated with a broken gauge symmetry.  

\subsubsection*{$U(1)^2$ Couplings}
\label{Sec: Charges}

If we choose to gauge two $U(1)$'s, it is possible that the Standard Model 
fermions are charged under both of them.  The charges of the light fermions 
under these $U(1)$'s, $Q^f_{1,2}$, can be written as
\begin{eqnarray}
Q^f_1 = \half(1 + R) Q^f_{\text{SM}}
\hspace{0.6in}
Q^f_2 = \half(1 -  R) Q^f_{\text{SM}},
\end{eqnarray}
where $R$ is the same for all fermions in a generation in order for anomalies 
to cancel.  The coupling of the Standard Model fermions to the $B'$ is then 
given by:
\begin{eqnarray}
\LL_{\text{Int}} = 
 g'(\cot 2 \theta'  + R \csc 2 \theta' ) \; B'_\mu\, j_{\text{F}}^\mu
\end{eqnarray}
where $j_{\text{F}}^\mu$ is the $U(1)_Y$ electroweak current.
Taking $R= 0$, in which case  the fermions are charged equally under
the two $U(1)$'s, the coupling to the $B'$ vanishes as 
$\theta' \rightarrow \frac{\pi}{4}$, and in the same limit the $B'$ coupling 
to the Higgs current also disappears.  This is the second half of the near 
oblique limit.  We will take $R=0$ throughout the paper.  

The Yukawa couplings of the first two generations to the Higgs
can be written in terms of $\Sigma$.  
For a Type I model, we might have
\begin{eqnarray}
\nonumber
\LL^{\text{Type I}}_{\text{Yuk}} &=& 
y_u f \, q^\alpha\, \Sigma_{\alpha 5} (\det \Sigma^*)^\half\, u^c  
+ y_d f \, q^\alpha\, \Sigma^*_{\alpha 5} (\det \Sigma)^\half\, d^c  
+ y_e f\,  l^\alpha\, \Sigma^*_{\alpha 5} \det(\Sigma)^\half\, e^c \\
&=& y_u\, q h_1 u^c + y_d\, q h_1^* d^c + y_e\, l h_1^* e^c +\cdots,
\end{eqnarray}
and for a Type II model,
\begin{eqnarray}
\nonumber
\LL^{\text{Type II}}_{\text{Yuk}} &=& 
y_u f \, q^\alpha\, \Sigma_{\alpha 5} (\det \Sigma^*)^\half\, u^c  
+ y_d f \, q^\alpha\, \Sigma_{\alpha 6} (\det \Sigma^*)^\half\, d^c  
+ y_e f\,  l^\alpha\, \Sigma_{\alpha 6} \det(\Sigma^*)^\half\, e^c \\
&=& y_u\, q h_1 u^c + y_d\, q h_2^* d^c + y_e\, l h_2^* e^c +\cdots .
\end{eqnarray}
These fractional powers of the non-linear sigma model field
can appear naturally in theories of strong dynamics and often
do so with small coefficients.  This might be relevant for understanding the 
lightness of the first two generations of fermions.  

\subsection{Third Generation Fermions and the Top Yukawa}
\label{Sec: Top}

Even if a product of $U(1)$'s is gauged, the third generation will
typically only be charged under one $U(1)$ because of the need to 
covariantize the top coupling.  Because this charge assignment is different 
than for the first two generations, to have a Yukawa coupling between the 
first two generations and the third requires operators with different forms 
than those involving the first two generations alone.  For instance, Yukawa 
couplings linking the light generations with the third generation could come 
from 
\begin{eqnarray}
\LL_{\text{Yuk 3\,1,2}}
= 
y_{u\;3\,1,2}f\; q_3^\alpha 
\Sigma_{\alpha 5} (\det \Sigma)^{\frac{2}{3}}  u_{1,2}^c 
+ y_{u\;1,2\,3}f\; q_{1,2}^\alpha 
\Sigma_{\alpha 5} (\det \Sigma)^{-\frac{1}{6}}  u_{3}^c 
+\cdots.
\end{eqnarray}
One might speculate that suppressions associated with the fractional powers 
of the non-linear sigma field contribute the hierarchical structure of the 
CKM matrix. 

Having the third generation charged differently than the first two generations
leads to FCNCs mediated by the $B'$.   Taking 
$\theta' = \frac{\pi}{4} +\delta \theta'$, the $B'$ coupling has a 
generational structure $g' \diag(2\delta \theta',2\delta \theta',1)$.  
After going to the mass eigenstate basis there will be off-diagonal entries
in this coupling matrix.  This leads to constraints
on the structure of the Yukawa matrices.  Typically these effects will
be adequately small if the light-heavy generational mixing is
predominantly in the up sector and is small or absent in the down 
and lepton sectors.    Such a scenario could give rise to interesting third 
generation flavor physics coming from the top sector and should be explored 
in more depth. 

All little Higgs models stabilize the top's quadratic divergence by adding 
vector-like fermions at the TeV scale, but they differ in the number
of degrees of freedom added and in turn in the symmetries that
they preserve.  We use the notation
that fields transform under $SU(3)_C\times SU(2)_L\times U(1)_Y$ as
\begin{eqnarray}
\nonumber
&q \sim (\mathbf{3},\mathbf{2}, \mathbf{+\frac{1}{6}})
\hspace{0.4in}
q^c \sim (\mathbf{\bar{3}},\mathbf{2}, \mathbf{-\frac{1}{6}})
\\
\nonumber
&u \sim (\mathbf{3},\mathbf{1}, \mathbf{+\frac{2}{3}}) 
\hspace{0.4in}
u^c \sim (\mathbf{\bar{3}},\mathbf{1}, \mathbf{-\frac{2}{3}}) 
\\
&d \sim (\mathbf{3},\mathbf{1}, \mathbf{-\frac{1}{3}}) 
\hspace{0.4in}
d^c \sim (\mathbf{\bar{3}},\mathbf{1}, \mathbf{+\frac{1}{3}})  .
\end{eqnarray}
The charges of the fields under the $[SU(2)\times U(1)]^2$ are
implicitly defined by their couplings to the non-linear sigma model
fields.  Fields that have vector-like masses are labeled with
$\tilde{\psi}$, while the chiral Standard Model
third generation fields are labeled with $\psi_3$.   These
fermions will typically have anomalies under $U(1)^2$,
which we assume are canceled at high energies by additional fermions.

The top quark mixes with vector-like fermions, and this
mixing will induce FCNCs mediated by the $Z^0$. 
Because these FCNCs require mixing through the
heavy fermions, they are probably too small to be detected, although
a more thorough investigation is necessary to say anything definite. 

We will consider a top sector given by
%Consider a top sector that gives mass to the top through 
%mixing rather than through
% direct interactions with $\Sigma$.   A particularly elegant 
%sector is given by 
taking a full set of vector-like quarks that transform
as fundamentals  of $SU(6)$.  It is different than the top sector
of \cite{Low:2002ws}; its main advantages are than it has fewer
parameters
and gives finite radiative contributions to the Higgs mass. A
discussion of other possibilities for the top sector is given in the
appendix. A setup like this was considered in
\cite{Nelson:2003aj} in the context of a composite Higgs ultraviolet
completion for the Littlest Higgs.  In that theory the vector-like
fermions were composites of some underlying strongly interacting gauge
theory and mixed with the top quark after chiral symmetry breaking 
at $\Lambda \sim 10 \TeV$.   A similar ultraviolet completion
could be pursued in this model with a similar top sector
by adding fermions that transform under $SU(2)_L\times U(1)_Y$ as
\begin{eqnarray}
X =
\left(\begin{array}{c}
\tilde{q}_1 \\
\tilde{q}_2 \\
\tilde{u} \\
\tilde{d} 
\end{array}\right)
%%\sim \left( \begin{array}{c} (\mathbf{3_c},\mathbf{2_{+\frac{1}{6}}})\\ (\mathbf{3_c},\mathbf{2_{+\frac{1}{6}}})\\ (\mathbf{3_c},\mathbf{1_{+\frac{2}{3}}})\\ (\mathbf{3_c},\mathbf{1_{-\frac{1}{3}}}) \end{array}\right)
\hspace{0.3in}
X^c =
\left( \begin{array}{c}
\tilde{q}_2{}^c \\
\tilde{q}_1{}^c \\
\tilde{d}^c \\
\tilde{u}^c 
\end{array}\right)
%%\sim \left( \begin{array}{c} (\mathbf{\bar{3}_c},\mathbf{2_{-\frac{1}{6}}})\\ (\mathbf{\bar{3}_c},\mathbf{2_{-\frac{1}{6}}})\\ (\mathbf{\bar{3}_c},\mathbf{1_{-\frac{2}{3}}}) \\ (\mathbf{\bar{3}_c},\mathbf{1_{+\frac{1}{3}}}) \end{array}\right).
\end{eqnarray}
These couple to $\Sigma$ with the top quark, $q_3$ and $u_3^c$, through
\begin{eqnarray}
\LL_{\text{top}} =
y_1 f 
\; X^c\, \Sigma\, X
+ y_2 f\; \tilde{u} u_3^c + 
\tilde{y}_2 f\; \tilde{q}^c_1 q_3 +\hc
\end{eqnarray}
There are other possible mixing terms such as $\det \Sigma\, \tilde{d} d_3^c$
and $\det \Sigma\, \tilde{q}^c_2 q_3$ that could be considered.  We
ignore these for simplicity, but if present they will affect the discussion 
of third generation physics in Sec. \ref{Sec: 3rd Gen}.
Notice that $y_1$ preserves both $SU(2)_R$ and $SU(2)_H$,
while $y_2$ breaks $SU(2)_R$ and $\tilde{y_2}$ breaks
$SU(2)_H$.

The quark singlet and doublet mix with the 
fermions and can be diagonalized with mixing angles 
$\vartheta_U$ and $\vartheta_Q$ respectively:
\begin{eqnarray}
\tan \vartheta_U = \frac{y_1}{y_2}
\hspace{0.3in}
\tan \vartheta_Q = \frac{y_1}{\tilde{y}_2}
\hspace{0.3in}
y_{\text{top}}^{-2} = 
|y_1|^2 \left(|y_1|^{-2} + |y_2|^{-2}\right)
\left(|y_1|^{-2} + |\tilde{y}_2|^{-2}\right).
\end{eqnarray}
The mass of the top quark is given by
\begin{eqnarray}
m_{\text{top}} = y_{\text{top}} v \cos \beta \simeq 175 \GeV .
\end{eqnarray}
The Yukawa couplings can be expressed in terms of the mixing angles
and top Yukawa coupling as
\begin{eqnarray} 
y_1 = \frac{y_{\text{top}} }{\cos\vartheta_U \cos\vartheta_Q}
\hspace{0.5in}
y_2 = \frac{y_{\text{top}} }{\sin\vartheta_U \cos\vartheta_Q}
\hspace{0.7in}
\tilde{y}_2 = \frac{y_{\text{top}} }{\cos\vartheta_U \sin\vartheta_Q},
\end{eqnarray}
while the masses are
\begin{eqnarray} 
\nonumber
m_{u_H} = \frac{ 2 y_{\text{top}} f }{\sin 2\vartheta_U \cos \vartheta_Q}
\hspace{0.6in}
m_{q_H} = \frac{ 2 y_{\text{top}} f }{\sin 2\vartheta_Q \cos \vartheta_U}\\
m_{d_H} = \frac{ y_{\text{top}} f }{\cos \vartheta_Q \cos \vartheta_U}
\hspace{0.6in}
m_{q'_H} = \frac{ y_{\text{top}} f }{\cos \vartheta_Q \cos \vartheta_U} .
\end{eqnarray}
The top-sector radiative correction to the Higgs mass is minimized for 
$\cos^2 \vartheta_U = \cos^2 \vartheta_Q = \frac{2}{3}$, which gives
\begin{eqnarray}
\nonumber
m_{u_H} = m_{q_H}= m_{\text{min}} \equiv
\frac{ 3 \sqrt{3} m_{\text{top}} }{ 2 \cos \beta} \frac{f}{v}
\hspace{0.6in}
m_{d_H} = m_{q'_H}= \frac{m_{\text{min}}}{\sqrt{3}}=
\frac{ 3  m_{\text{top}} }{ 2 \cos \beta} \frac{f}{v} .
\end{eqnarray}

\subsubsection*{Radiative Corrections}

The top Yukawa preserves an $SU(6)_L \times SU(6)_R$ chiral symmetry, 
preventing a quadratically divergent mass term for the Higgs from being 
generated at one or two loops.  Therefore the finite one loop contribution
to the mass dominates and is calculable.
The one loop contribution from the Coleman-Weinberg potential is
\begin{eqnarray}
\label{Eq: SU6 Rad Cor}
V_\eff =
- \frac{3 y^2_{\text{top}}}{8\pi^2}
\left(m_{u_H}^{-2}- m_{q_H}^{-2}\right)^{-1}
\log \frac{m_{q_H}^2}{m_{u_H}^2} \; |h_1|^2 
%+ \lambda_{11}' |h_1|^4 
+ \cdots.
\end{eqnarray}
Taking $m_{u_H}\rightarrow m_{q_H}\rightarrow m$, 
this becomes
\begin{eqnarray}
\nonumber
V_{\eff} &=& - \frac{3 m^2_{\text{top}}}{8\pi^2 v^2\cos^2\beta} 
m^2\; |h_1|^2,
%\\
%&=&-\frac{81}{32 \pi^2} \left(\frac{ m_{\text{top}}}{v \cos
%    \beta}\right)^4 f^2 |h_1|^2.
\end{eqnarray}
and further taking $m\rightarrow m_{\text{min}}$ gives the minimum
negative contribution to the Higgs mass squared,
\begin{equation}
\delta m^2 =\frac{81}{32 \pi^2} \left(\frac{ m_{\text{top}}}{v \cos
    \beta}\right)^4 f^2.
\label{Eq: SU6 Rad Cor2}
\end{equation}
Notice that this radiative correction is sensitive to $\beta$.
This is the only asymmetry between the $h_1$ and $h_2$ masses generated at
one loop.  

\subsection{Electroweak Symmetry Breaking}
\label{Sec: EWSB}

At this point we can consider electroweak symmetry breaking.
The Higgs doublets are classically massless but pick up 
$\mathcal{O}(v^2)$ masses from radiative corrections
to the tree-level Lagrangian.  The gauge and scalar contributions 
to the little Higgs masses are positive while the fermions give a negative 
contribution.  The  Peccei-Quinn symmetry
forbids the $b$ term $b h_1^\dagger h_2$, 
necessary for viable electroweak symmetry
breaking.   There are a number of possible operators that could be
generated in the ultraviolet to break this symmetry.  For instance, 
in \cite{Low:2002ws} the operator
\begin{eqnarray}
\OO_{\text{$b$-term}} = 
b f^2  \epsilon^{\alpha \beta}\Sigma_{\alpha 5}\Sigma_{\beta \alpha'}
\Sigma^*{}^{\alpha' 5} +\hc .
\end{eqnarray}
was suggested for this purpose.
%In the top sector used by \cite{Low:2002ws} 
%this symmetry was broken and top physics generated  $h_1^\dagger h_2$.
Here we take a phenomenological approach and 
simply write down a $b$ term with an appropriate coefficient.
%There are several such operators that could work that can
%be generated by additional TeV to multi-TeV physics 
%that breaks this symmetry.
The potential for the Higgs doublets is  then
\begin{equation}
%\nonumber
V_{\text{eff}} = \left(\begin{array}{cc} h_1^\dagger & h_2^\dagger\end{array}\right)  
\left(\begin{array}{cc} m^2_1 & b \\ b^* & m^2_2\end{array}\right)
\left(\begin{array}{c} h_1 \\h_2\end{array}\right)\\+ \lambda |h_1^\dagger h_2|^2.
\end{equation}
The phase of $b$ can be rotated away with a $U(1)_{PQ}$ transformation.

We should mention that there are also additional, subleading contributions to the quartic potential
($|h_1|^4$, $|h_1|^2 |h_2|^2$, and $|h_2|^4$ terms), that
come from logarithmic running from the cutoff to the weak scale.
These terms can be regarded as perturbations on top of the Higgs potential as
far as the Higgs spectrum is concerned. However, they induce \begin{it}one-loop\end{it}
quadratic divergences in $\delta m^2$.  For example, the top sector induces an $SU(2)_{H}$ -- violating
quartic term $|h_1|^4$ with a coefficient of $\OO(0.1)$, and we estimate that this term 
gives a contribution to $\delta m^2$ that is $\OO(20\%)$ of the original top contribution 
(the naive expectation  is that the signs are opposite). This is the largest source of uncertainty for
$\delta m^2$.

To have stable electroweak symmetry breaking the following conditions
in the mass squared matrix must be met:
\begin{eqnarray}
m_1^2 > 0 
\hspace{0.5in}
m_2^2 >0
\hspace{0.5in}
m_1 m_2 - b < 0 .
\end{eqnarray}
The mass terms $m^2_{1,2}$ are generated radiatively
\begin{eqnarray}
\label{Eq: Radiative Masses}
m_1^2 = \Delta m^2 - \delta m^2 \hspace{0.5in}
m_2^2 = \Delta m^2,
\end{eqnarray}
where $\Delta m^2$ arises from the gauge and scalar 
sectors and is logarithmically enhanced while $\delta m^2$ comes
from the top sector.
The vacuum expectation values of the doublets take the form
\begin{eqnarray}
\langle h_1 \rangle =\frac{1}{\sqrt{2}} \left( \begin{array}{c} 0\\ v \cos \beta \end{array}\right)
\hspace{0.2in}
\langle h_2 \rangle = \frac{1}{\sqrt{2}}\left( \begin{array}{c} 0\\ v\sin \beta \end{array}\right).
\end{eqnarray}

An important parameter in the Higgs sector is $\tan \beta$, not only
because it affects the top's radiative corrections
through  Eq.~(\ref{Eq: SU6 Rad Cor2}), but also
because deviations from $\tan \beta = 1$ will contribute to
oblique corrections.  We find
\begin{equation}
\tan^2\beta=\frac{m_1^2}{m_2^2}=\frac{\Delta m^2 -\delta m^2}{\Delta m^2}.
\label{Eq: tanbeta}
\end{equation}
Thus we have $\tan\beta<1$, and it is quite plausible to have $\tan
\beta$ near unity.
The other electroweak symmetry breaking parameters can be calculated
in terms of the masses and quartic coupling,
\begin{eqnarray}
\frac{1}{2} \lambda v^2 &=& m^2_{A^0} - m^2_{H^\pm} \label{eq:ewsb}\\
\label{Eq: Alpha}
\tan 2 \alpha &=& (1-2 x) \tan 2 \beta.
\end{eqnarray}
with $\alpha$ being the mixing angle for the $h^0 - H^0$ sector.
%The ratio of the charged Higgs mass to the pseudo-scalar Higgs
%mass plays an important role in this theory and is given by
Here we have introduced the parameter
\begin{eqnarray}
x \equiv \frac{m^2_{H^\pm}}{m^2_{A^0}}=\frac{m_1 m_2}{b}. 
\hspace{0.7in} 0 \le x \le 1 .
\end{eqnarray}
This parameter plays an important role in the theory because it 
is connected to the fine-tuning of the Higgs potential: the closer
$x$ is to unity, the more finely tuned the theory is.   A more
convenient fine tuning parameter is given by
\begin{eqnarray}
\kappa = x^{-1} -1  \equiv \frac{\lambda v^2}{2 m^2_{H^\pm}} 
\end{eqnarray}
From Eq.~(\ref{eq:ewsb}) one can see that this measure of fine tuning
is a reasonable one. 

The masses of the five physical Higgs particles are
\begin{eqnarray}
\nonumber
m^2_{H^\pm} &=& m_1^2 + m_2^2 = x \; m^2_{A^0}\\ 
\nonumber
m^2_{A^0} &=& \frac{2 b}{\sin 2 \beta} = m^2_{H^\pm}+ \half \lambda v^2\\
\nonumber
m^2_{h^0} &=&  m^2_{A^0}\frac{\left( 
1 - \sqrt{\cos^2 2\beta +(1-2 x)^2 \sin^2 2 \beta} 
\right)}{2}\\
m^2_{H^0} &=& m^2_{A^0} - m^2_{h^0}.
\label{Eq: Higgs Masses}
\end{eqnarray}
The heaviest Higgs is the pseudo-scalar $A^0$, a fact
which will have consequences for precision electroweak observables.

%%%%

\section{Precision Electroweak Observables}
\label{Sec: PEW}
\setcounter{equation}{0}
\renewcommand{\theequation}{\thesection.\arabic{equation}}

In this section the effects of the $SU(6)/Sp(6)$ model
on precision electroweak observables are  discussed.  
%There are several types of effects to calculate.   
First let us estimate the typical size of incalculable
cutoff-sensitive contributions.
These theories have a low cut-off of $4\pi f \sim 10$~TeV, and
higher order terms in the Lagrangian such as those mentioned
in Eq. \ref{Eq: NLSM Kin 4D} appear as
\begin{eqnarray}
\LL \supset \frac{1}{\Lambda^2} \OO_{6} + \cdots .
\end{eqnarray}
Precision electroweak constraints are essentially
constraints on the coefficients of dimension 6  operators.  For instance
the $S$ parameter is generated by $h^\dagger W_{\mu\nu}h B^{\mu\nu}$, which 
gives a  contribution to $\Delta S$ 
\begin{eqnarray}
\alpha \Delta S \simeq g g'\frac{v^2}{\Lambda^2} \simeq
\frac{gg'}{(4\pi)^2} \frac{v^2}{f^2}. 
\end{eqnarray}
This means that this operators leads to an intrinsic cut-off uncertainty of
$\Delta S \simeq \pm 0.02$. This is parametrically the same
size as would be given by dimension 8 operators suppressed by $f^{-4}$, 
and therefore it is typically not necessary to go beyond calculating dimension
6 operators in little Higgs models without making assumptions
about cut-off scale physics.   However because $v/f$ in this model
is small, we check the dimension 8 operators to verify this intuition.

In section \ref{Sec: PEWCurrents}, we 
%discuss the near-oblique limit that is necessitated
%by precision electroweak constraints on 
examine non-oblique corrections from $Z^0$-pole observables 
and four-Fermi operators and consider the near-oblique limit.
These correction come from integrating out the $W'$
and place limits on the mixing angle $\theta$,
related to the ratio of the $SU(2)$ gauge couplings.  
If we choose to cut off the $U(1)_Y$ quadratic divergence with
the $B'$, there are constraints on $\theta'$  as well.  
%There is a limit where all non-oblique corrections 
%related to the first
%two generations disappear and this is called the ``near-oblique
%limit'' of the model.  

In Sec.~\ref{Sec: Oblique} oblique corrections are computed.   
At tree level, these include contributions mediated by
the gauge bosons as well as contributions from higher order
terms in the non-linear sigma model kinetic term.  We also calculate
radiative effects from the two Higgs doublets and the top
sector.  These radiative effects are  important to consider 
when constraining the model.

The non-oblique corrections that cannot be removed by the
near-oblique limit involve the third generation, and are considered 
in Sec.~\ref{Sec: 3rd Gen}. The
most important involve mixing of the left-handed bottom quark. We do
not perform a careful analysis of these effects, but speculate that   
these corrections may offer explanations for possible anomalies
in $Z^0$-pole bottom physics.  

Another serious constraint, although not technically a precision
electroweak observable, comes from  direct production of the $B'$
if two $U(1)$'s are gauged.  Sec.~\ref{Sec: B' Production}
provides a brief discussion of the rather subtle collider
physics involving the $B'$ in this model.

\subsection{Electroweak Currents and Four-Fermi Operators -- The Near 
Oblique Limit}
\label{Sec: PEWCurrents}

We begin by considering
the modifications to electroweak currents and their effect
on four-Fermi operators at low energies.  In this section
we will only look at the effects on the first two
generations and consider the third generation separately in
Sec. \ref{Sec: 3rd Gen}.
The $W'$ couples to the Goldstone bosons 
through the current interaction of the Higgs, $W'{}^a j_{W' H}^a$, where the 
current is given by 
\begin{eqnarray}
\nonumber
j_{W'\,H}^a{}_\mu &=&
g \cot 2 \theta [i h_1^\dagger \sigma^a  \overleftrightarrow{D}_\mu h_1
+ i h_2^\dagger \sigma^a  \overleftrightarrow{D}_\mu h_2]
= \frac{i g \cot 2\theta v^2}{2} 
 \Tr \sigma^a \omega \overleftrightarrow{D}_\mu \omega^\dagger
+\cdots\\
&=&   i g \cot 2 \theta j_{\omega\;\mu}^a +\cdots
\end{eqnarray}
Here the $\cdots$ represents interactions involving the physical
Higgs bosons that are unimportant for precision electroweak physics.
The the Higgs $B'$ current is much the same:
\begin{eqnarray}
\nonumber
j_{B'\,H}^\mu &=&
\frac{i}{2} h_1^\dagger  \overleftrightarrow{D}^\mu h_1
+ \frac{i}{2} h_2^\dagger   \overleftrightarrow{D}^\mu h_2
= \frac{i v^2}{4}
\Tr \sigma^3 \omega^\dagger \overleftrightarrow{D}_\mu \omega
+\cdots\\
&=&  j_{\omega\;\mu} +\cdots .
\end{eqnarray}

There are then Higgs-fermion interactions mediated by
the $W'$ and $B'$, which directly modify $Z^0$-pole observables:
\begin{eqnarray}
\nonumber
\LL_{\text{H F}} &=& 
\frac{c_{HF}^L}{v^2} j_{\omega\;\mu}^a{}\; j^\mu{}^a{}_{\text{F}} 
+\frac{c_{HF}^Y}{v^2} j_{\omega\;\mu}\; j^\mu{}^a{}_{ \text{F}}+\cdots\\
\nonumber
&=&\frac{j_\mu^a{}_{W' \text{H}}\; j^\mu{}^a{}_{W' \text{F}}}{2m^2_{W'}}
+
\frac{j_\mu{}_{B' \text{H}}\; j^\mu{}_{B'\text{F}}}{2m^2_{B'}}\\
&=&   \frac{\sin^2 \theta \cos 2\theta}{2 f^2} 
j_{\omega}^{a\,\mu} j_{\text{F}}{}_{a\,\mu} 
+   \frac{\cos 2 \theta'(\cos 2\theta' +R) }{ \bar{f}^2} 
j_{\omega}^{\mu} j_{\text{F}}{}_{\mu}. 
\end{eqnarray}
So the coefficients of the dimension 6 operators are
\begin{eqnarray}
\label{Eq: cHF}
c_{HF}^L = 2 \sin^2 \theta \cos 2\theta \frac{v^2}{ f^2} 
\hspace{0.5in}
c_{HF}^Y = \cos 2\theta'(\cos 2\theta' + R)  \frac{v^2}{\bar{f}^2},
\end{eqnarray}
where we have included the possibility of $R \ne 0$, in the notation
of Sec.\ref{Sec: Charges}.  The
$c_{HF}^L$ operator can be rewritten as
\begin{eqnarray}
c_{HF}^L = 2 (1 - \tan ^2 \theta) \frac{m^2_{W^\pm}}{m^2_{W'}}. 
%\lsim \delta_{c_{HF}^L}
\end{eqnarray} 
It requires a full fit to know the limits on these interactions,
but to a good approximation they are not problematic if they are
suppressed by roughly $4 \TeV$ \cite{Barbieri:1999tm}.
This translates into $c_{HF} \lsim 1/250$ (note that in the $SU(2)$ currents we use $\Tr \sigma^a \sigma ^b = \half
\delta^{ab}$ and the $U(1)$ fermion currents contain charges, so
our normalizations differ from those used in \cite{Barbieri:1999tm}).
Taking $\theta' = \frac{\pi}{4}+ \delta \theta'$ and $R=0$,
the constraints from the operators involving the $B'$ reduce to
\begin{eqnarray}
\label{Eq: theta' NOL}
\delta \theta' \lsim \frac{1}{10} \; \frac{\bar{f}}{2 \TeV},
\end{eqnarray}
while the constraints from the $W'$ reduce in the small $\theta$ limit
to a constraint on the mass of the $W'$,
\begin{eqnarray}
m_{W'} \gsim 1.8 \TeV .
\end{eqnarray}
According to Eq.~(\ref{eq:wprimemass}), this  means that we have  
$\sin \theta \lsim \frac{1}{5}$  for $f = 700 \GeV$, which translates into
$g_2 \gsim 3.0$, so there is still room for the coupling to be perturbative.  

The second modification of fermion interactions are
four-Fermi interactions that are constrained by both
low energy physics such as  $G_F$ and atomic parity violation, and  high
energy tests of fermion compositeness:
\begin{eqnarray}
\nonumber
\LL_{\text{F F}} &=&
\frac{c_{FF}^L}{v^2}  j^\mu{}^a{}_{\text{F}} j_\mu{}_a{}_{\text{F}} 
+\frac{c_{FF}^Y}{v^2} j^\mu{}_{ \text{F}} j^\mu{}_{ \text{F}}
\\
\nonumber
&=&-\frac{(j_\mu^a{}_{W'\text{F}})^2}{2M^2_{W'}} +
-\frac{(j_\mu{}_{B'\text{F}})^2}{2M^2_{B'}}\\
&=&
-\frac{\sin^4\theta}{ f^2}
j_{\text{F}}{}^{a\,\mu} j_{\text{F}}{}_{a\,\mu}
-\frac{2 (\cos 2\theta' +R)^2}{ \bar{f}^2}
j_{\text{F}}{}^{\mu} j_{\text{F}}{}_{\mu}.
\end{eqnarray}
The coefficients for the dimension 6 operators are
\begin{eqnarray}
\label{Eq: cFF}
c_{FF}^L = - \sin^4 \theta \frac{v^2}{f^2}
\hspace{0.5in}
c_{FF}^Y = - (\cos 2\theta' + R)^2  \frac{v^2}{\bar{f}^2 } .
\end{eqnarray}
The sensitivity to these terms is generally subdominant to the sensitivity to 
the operators that modify $Z^0$ pole observables.

In summary,  as $\theta \rightarrow 0$ and as 
$\theta' \rightarrow \frac{\pi}{4}$, the non-oblique corrections
vanish for the first two generations.  In the next section we calculate
contributions to the $S$ and $T$ parameters near this limit.  

\subsection{Oblique Corrections}
\label{Sec: Oblique}

Precision electroweak tests impose stringent constraints on custodial $SU(2)$ 
violation, modifications of the interactions of the Goldstone bosons eaten by 
the $W^\pm$ and $Z^0$.  These Goldstone bosons live inside the Higgs doublets as
\begin{eqnarray}
h_1(x) = \frac{v \cos\beta}{\sqrt{2}}\; 
\omega(x) \left(\begin{array}{c}0\\1\end{array}\right)  
+ \cdots
\hspace{0.5in}
h_2(x) = \frac{v \sin \beta}{\sqrt{2}}\; 
\omega(x) \left(\begin{array}{c}0\\1\end{array}\right) 
+ \cdots .
\end{eqnarray}
When written in terms of the electroweak chiral Lagrangian, violations 
of $SU(2)_C$ stem from the higher order interaction
\begin{eqnarray}
\label{Eq: EW T}
\OO_T = c_T v^2 \big(\Tr \sigma^3 \omega^\dagger D_\mu \omega\big)^2 +\hc
\Rightarrow  \alpha \Delta T = + c_T
\label{eq:rho}
\end{eqnarray}
In this section we calculate the coefficient of this operator from
various sources.  

Typically there are four new sources for $SU(2)_C$ violation
in little Higgs models, which we will discuss in turn.  
First we consider the effects of integrating out the heavy 
$W'$ and $B'$ gauge bosons. Next, we calculate the $SU(2)_C$ violation 
coming from the non-linear sigma model
structure itself.  Finally, we analyze the radiative corrections of the Higgs
doublets and the top partners.  

\subsubsection{Gauge Bosons}

The most straightforward oblique corrections come from integrating out the 
new gauge bosons.  The leading Lagrangian for both the $W'$ and $B'$,  
including the current interaction in Eq. \ref{Eq: Current Interactions}, is 
\begin{eqnarray}
\LL =  
- \frac{1}{4}W'{}_{\mu\nu}^a{}^2 + \frac{m^2_{W'}}{2} W'{}_\mu^a{}^2
+ W'{}^a_\mu j_{a\,H}^\mu 
- \frac{1}{4}B'{}_{\mu\nu}{}^2 + \frac{m^2_{B'}}{2}B'{}_\mu{}^2
+ B'_\mu j_{H}^\mu .
\end{eqnarray}
The source term can be eliminated by shifting the gauge bosons,
producing an effective action 
\begin{eqnarray}
\LL_{\eff} &= &
- \frac{(j^{a}_{H\,\mu})^2}{2 m_{W'}^2} 
- \frac{(j_H^\mu)^2}{2 m_{B'}^2} 
-  \frac{(D_{[\mu}j^{a}_{H\, \nu]})^2}{4 m_{W'}^4}
- \frac{(D_{[\mu}j_{H\,\nu]})^2}{4 m_{B'}^4} +\cdots
\label{Eq: Oblique Gauge}
\end{eqnarray}
The first term simply renormalizes $v$ by a finite amount and
therefore is not important. When expanded, the second term
gives operators of the form
\begin{eqnarray}
\LL= -  \frac{ \cos^2 2 \theta'}{2\bar{f}^2} 
[h_1^\dagger D h_1 + h_2^\dagger D h_2]^2
+\hc
\end{eqnarray}
This operator violates $SU(2)_C$ and gives a contribution to  $T$
\begin{eqnarray}
\label{Eq: T B'}
\alpha \Delta T =  \frac{ v^2}{2 \bar{f}^2} \cos^2 2\theta',
\end{eqnarray}
Taking $\theta' =  \frac{\pi}{4} + \delta \theta'$, this becomes 
\begin{eqnarray}
\Delta T_{B'} \simeq 4 \,\delta \theta'{}^2 \frac{ (2 \TeV)^2}{ \bar{f}^2}.
\end{eqnarray}
Using the reference value $\bar{f} \simeq 2$~TeV, this contribution 
is small when the non-oblique corrections from the $B'$ are adequately suppressed.
If $\theta'$ satisfies Eq. \ref{Eq: theta' NOL} we get 
$\Delta T_{B'} \lsim +0.04$.

The last two terms in Eq. \ref{Eq: Oblique Gauge} give
contributions to the $S$ and $U$ parameters, of order $v^4/f^4$:  
\begin{eqnarray}
\nonumber
\alpha \Delta S &=&
-\sin^2 \theta_w \cos^2 2\theta \sin^2 2\theta \frac{v^4}{4 f^4}
- \sin^2 \theta_w \cos^2 2\theta' \sin^2 2\theta' \frac{v^4}{4 \bar{f}^4},\\
\alpha \Delta U &= &
+ \sin^2 \theta_w \cos^2 2\theta' \sin^2 2\theta' \frac{v^4}{4 \bar{f}^4}.
\end{eqnarray}
Again using the reference values 
$f \sim 700 \GeV$, $\bar{f} \sim 2\TeV$, $\theta = \delta \theta$, and 
$\theta' = \frac{\pi}{4} + \delta \theta'$, these become
\begin{eqnarray}
\Delta S = -\half \delta \theta^2 - \OO(10^{-3}) \delta \theta'{}^2 
\hspace{0.5in}
\Delta U =  + \OO(10^{-3}) \delta \theta'{}^2. 
\end{eqnarray}
For $\delta \theta \lsim \frac{1}{5}$, required to adequately suppress the non-oblique corrections
from the $W'$,  the contribution  $\Delta S \simeq -0.02$
is of similar size to the cut-off uncertainty.   
There are also operators of dimension 8 that contribute to $T$ that
make small corrections to Eq. \ref{Eq: T B'}.

\subsubsection{Non-Linear Sigma Model}
\label{Sec: NLSM}

The non-linear sigma model structure leads to  $SU(2)_C$ violating
operators obtained by expanding the kinetic term to quartic order.  
Expressing these operators
in terms of the Higgs doublets we find
\begin{eqnarray}
\LL = \frac{1}{2 f^2} | h_1^\dagger D h_1 + h_2^\dagger D h_2|^2
+ \frac{1}{2 f^2} |h_1 D h_2 + h_2 D h_1|^2.  
\end{eqnarray}
The first term gives a mass to the $Z^0$ while the second, which contracts the Higgs doublets
with the epsilon tensor, gives a mass to the $W^\pm$.
These terms have the  property that at $\tan \beta =1$ there is no $SU(2)_C$
violation,
\begin{eqnarray}
\alpha \Delta T = \frac{v^2}{4 f^2}  \cos^2 2\beta.
\label{Eq: NLSM}
\end{eqnarray}
If we require this single contribution to be 
$\Delta T_{\text{nl$\sigma$m}} \lsim 0.2$,
then the following limit is  obtained:
\begin{eqnarray}
|\cos 2 \beta| \le 0.2\; \frac{f}{700 \GeV}.
\label{Eq: Tan Beta Limit}
\end{eqnarray}
One should keep in mind, however, that  additional $\tan \beta$ 
dependence enters in oblique corrections from the Higgs doublets and 
top sector, so it is not appropriate to impose this bound strictly.
For the reference values given in  Sec.~\ref{Sec: Overview}.
one finds $\cos 2\beta \simeq 0.18$, giving $\Delta T_{\text{nl$\sigma$m}} \simeq 0.12$.

\subsubsection{Two Higgs Doublets}

The $T$ parameter also receives a one-loop contribution from 
the Higgs bosons.  It is known that this contribution can
be either positive or negative.  Typically it is positive if
the $H^\pm$ states are either lighter or heavier than all the
neutral states, and negative if there are neutral Higgs
states both lighter and heavier than it.   The Higgs potential of
this theory makes the pseudoscalar 
the heaviest Higgs boson, giving a negative contribution.  
\begin{eqnarray}
\nonumber
\Delta T_{\text{Higgs}} &=& 
\frac{1}{16 \pi\sin^2\theta_{\text{w}} m^2_{W^\pm}} \Big(
F(m^2_{A^0},m^2_{H^\pm})\\
\nonumber
&&
\hspace{0.5in} 
+ \cos^2(\alpha -\beta) \big(
F(m^2_{H^\pm}, m^2_{h^0}) -F(m^2_{A^0}, m^2_{h^0}) 
+ \widehat{T}_{\text{SM}}(m^2_{H^0})\big)\\
&&
\hspace{0.5in} 
+ \sin^2(\alpha -\beta) \big(
F(m^2_{H^\pm}, m^2_{H^0}) -F(m^2_{A^0}, m^2_{H^0}) 
+ \widehat{T}_{\text{SM}}(m^2_{h^0})\big)
\Big),
\end{eqnarray}
where
\begin{eqnarray}
F(x,y) &=& \half(x + y) - \frac{xy}{x -y} \log\frac{x}{y}\\
\nonumber
\widehat{T}_{\text{SM}}(m^2)&=&  F(m^2,m^2_{W^\pm}) - F(m^2,m^2_{Z^0})\\
&&+ \frac{4 m^2 m^2_{W^\pm}}{m^2 - m^2_{W^\pm}} \log \frac{m^2}{m^2_{W^\pm}}
- \frac{4 m^2 m^2_{Z^0}}{m^2 - m^2_{Z^0}} \log \frac{m^2}{m^2_{Z^0}}.
\end{eqnarray}

The Higgs doublets also give a contribution to the
$S$ parameter, 
\begin{eqnarray}
\nonumber
S &=& \frac{1}{12 \pi}\Big( 
\cos^2(\beta -\alpha) \log\frac{m^2_{H^0}}{m^2_{h^0}}
- \frac{11}{6} + \\
&&
\hspace{0.3in}
\sin^2(\beta -\alpha) G(m^2_{H^0}, m^2_{A^0}, m^2_{H^\pm})
+\cos^2(\beta -\alpha) G(m^2_{h^0}, m^2_{A^0}, m^2_{H^\pm})\Big),
\end{eqnarray}
where
\begin{eqnarray}
G(x,y,z) = \frac{x^2 + y^2}{(x^2 - y^2)^2}
+ \frac{(x- 3 y)x^2 \log\frac{x}{z} - (y - 3x) y^2 \log \frac{y}{z}}{(x-y)^3} .
\end{eqnarray}
The contribution to $S$ is positive.  The magnitudes of both $S$ and $T$
grow with the quartic coupling $\lambda$.  
The importance of the oblique corrections from the Higgs bosons 
will become evident in Sec. \ref{Sec: Results}.

\subsubsection{Top Sector}
\label{Sec: Top T}

Because they mix with the Standard Model chiral top, heavy fermions contribute 
to the $S$ and $T$ parameters at loop level. 
Parametrically, their contributions are suppressed by a factor of 
$v^2/f^2$ with respect to the Standard Model top contribution, but we find 
that these contributions are still significant.  

\begin{figure}
\begin{center}
\epsfig{figure=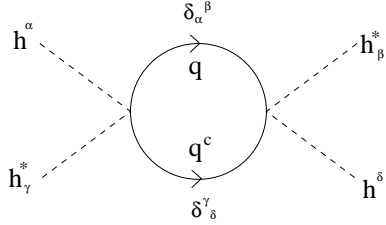,width=5cm} 
\caption{
Logarithmically divergent diagram to $T$ from quark doublets.
The contractions of the $SU(2)_L$ indices are shown.
\label{Fig: T Div}
}
\end{center}
\end{figure}

This theory contains non-renormalizable couplings, and the diagram shown in 
Fig.~\ref{Fig: T Div} can in principle contribute to $T$. When the heavy quark 
is an 
$SU(2)$ singlet, as in the Littlest Higgs model or in the SU(5) minimal top 
sector presented in the appendix, this diagram does not 
contribute to $T$. However, when there are heavy doublets as for the top sector
of Sec.~\ref{Sec: Top}, logarithmically divergent contributions to $T$ arise. 
This log divergence renormalizes the breaking scale $f$, as it can be shown 
that the log divergences sum into the operator
\begin{eqnarray}
\LL_{\text{Top Log}} =  - \frac{3 y^2_{\text{top}} f^2}{8 \pi^2}
\log \Lambda^2 \Tr D_\mu \Sigma^\dagger D^\mu \Sigma.
\end{eqnarray}
Because the non-linear sigma model self-interactions give a
tree-level contribution
to $T$ away from $\tan \beta = 1$ (Eq.~(\ref{Eq: NLSM})), 
this logarithmically divergent
renormalization of $f$ gives an additional 
contribution to $T$.  To deal with this
we simply absorb the log divergence into an effective breaking
scale and subtract the log divergence in the $\overline{ \text{MS}}$ 
scheme with subtraction scale of $1$ TeV.

Analytic results for the oblique 
corrections  from the top sector are not particularly enlightening
because they involve the
diagonalization of  $3\times 3$ and  $4\times 4$
matrices, without a good expansion parameter.
But we have calculated these corrections numerically, and we find that
there  are regions of parameter space where the top contribution to $T$ is 
acceptably small, and moreover, these regions 
largely include those that
%are essentially the same that
give the smallest values for 
the radiatively generated Higgs mass, $\delta m^2$. 
As a general feature, we find that as the mixing angles $\vartheta_Q$ and 
$\vartheta_U$ become small -- in which case the physical left and right-handed
top quarks $q_0$ and $u^c_0$ 
are primarily contained in the six-plet of fermions -- the contribution
to $T$ becomes small as well.  

\begin{figure}
\begin{center}
\epsfig{figure=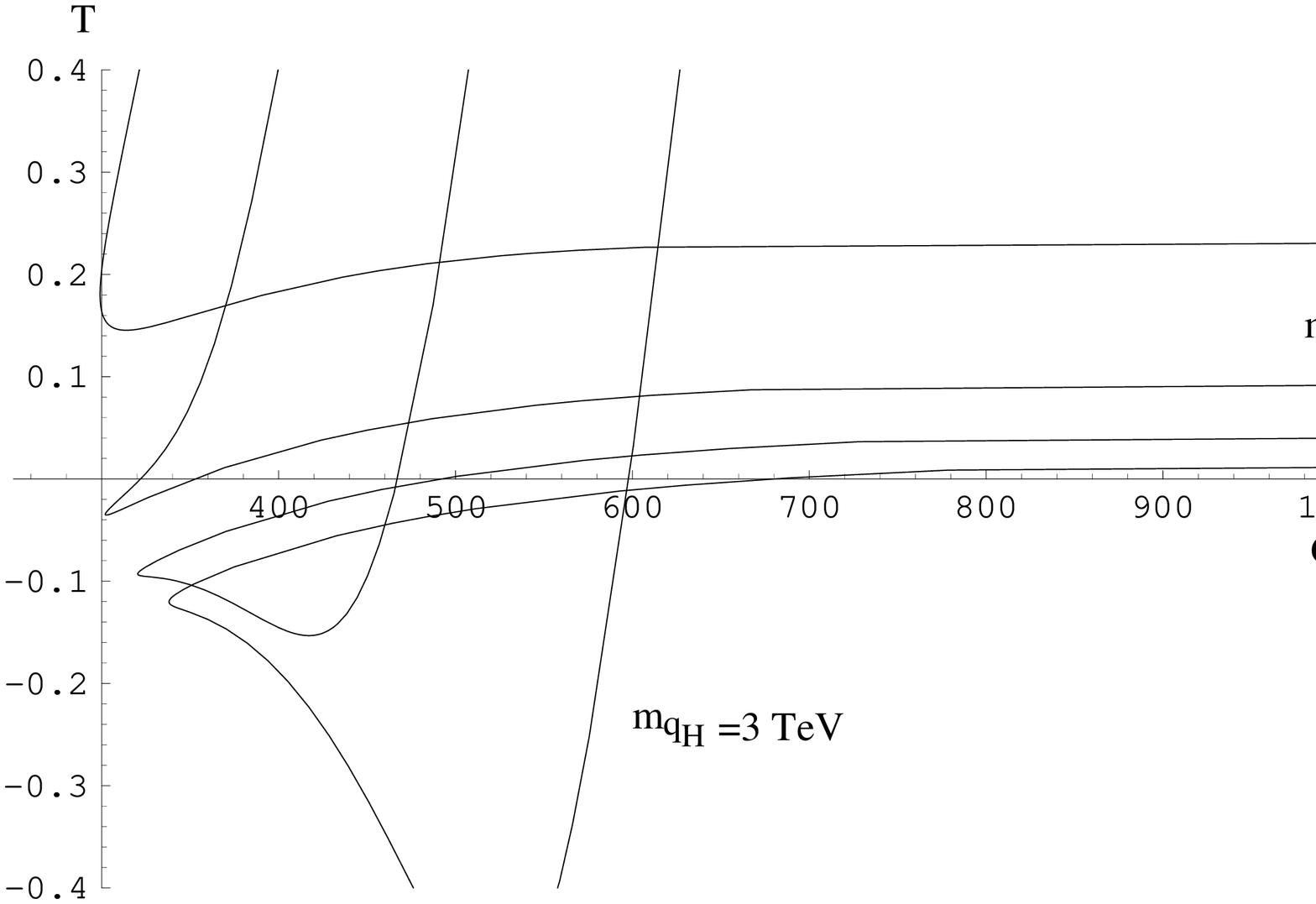,width=7cm} 
\epsfig{figure=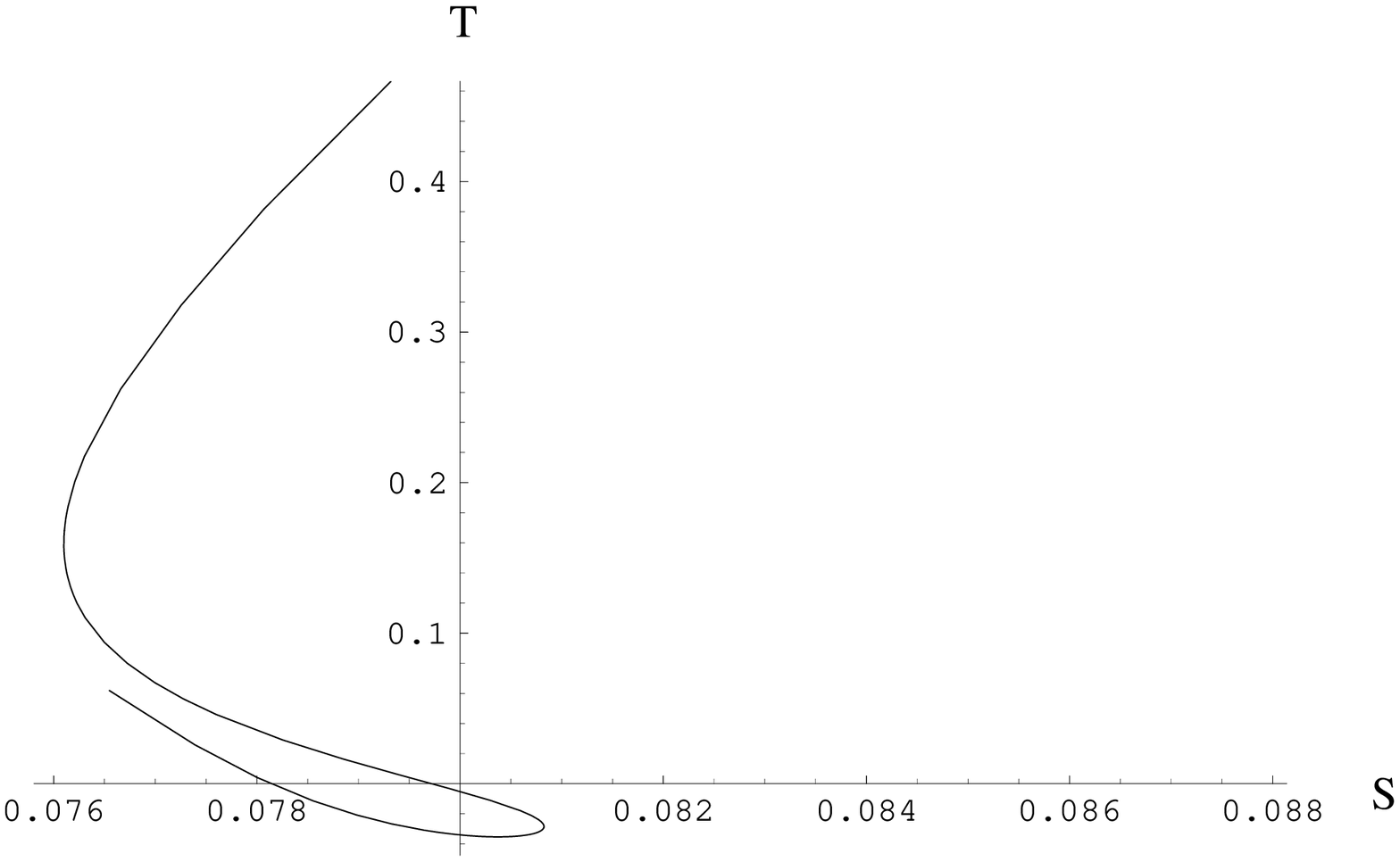,width=7cm}
\caption{
$T$ as a function of $\delta m$ for $f=700 \GeV$, $\tan \beta=0.8$
and various values of $m_{q_H}$ as a function of $\delta m$. The plot on
the right hand side shows T vs S for $f=700 \GeV$, $\tan \beta=0.8$ and
$m_{q_H} = 2 \TeV$ 
\label{Fig: T_sixplet}
}
\end{center}
\end{figure}

One way to fix all of the parameters of the top sector is
by specifying values for $\tan \beta$, $f$,
$\delta m^2$, and any one of the heavy quark masses.  After doing
so one can calculate the contribution to $T$.
In Fig.~\ref{Fig: T_sixplet}, we show $T$ vs. 
$\delta m \equiv \sqrt{\delta m^2}$ for 
$f=700 \GeV$, $\tan \beta=0.8$ and various values of $m_{q_H}$. 
The odd shapes of the contours simply reflect the fact that different
parameters can give the same $\delta m^2$.
We see that $\delta m$ is always   
larger than $300 \GeV$, consistent with the minimum value given in 
Eq.~\ref{Eq: Intro Top Rad} for $\tan \beta=0.8$. There is a sizable region 
of parameter space where the contribution to $T$ is acceptably small,
but note that both positive and negative contributions are possible.
In Fig.~\ref{Fig: T_sixplet} we also show a plot of $T$ vs. $S$ for 
$m_{q_H}=2$~TeV, which shows that the parameters that give adequately 
small $T$ give a positive contribution $\Delta S \simeq 0.08$.

\subsection{Third Generation Physics}
\label{Sec: 3rd Gen}

We have seen that there is a simple limit where non-oblique
corrections associated with the light two generations vanish.
This limit does not eliminate non-oblique corrections associated with the
third generation because (i) the third generation is assumed to be
charged under only a single $U(1)$, so it still couples significantly
to the $B'$ in the limit where the other generations decouple, and
(ii) third generation quarks mix with heavy fermions.
In the $SU(6)/Sp(6)$ model this mixing involves not just the top but
also the left-handed bottom quark, whose interactions are 
experimentally constrained.

At colliders, third generation physics has had a history of 
appearing anomalous, and mixing of the third generation fermions 
has in fact been proposed to resolve apparent anomalies \cite{Bamert:1996px}.  
There is presently no clear correct interpretation of precision 
data involving the third generation, and our approach has been simply to 
focus on the lighter generations and on oblique corrections.  
But the effects associated with the third generation may be 
important and deserve a more careful study than we give here.

\begin{figure}[ht]
\centering\epsfig{figure=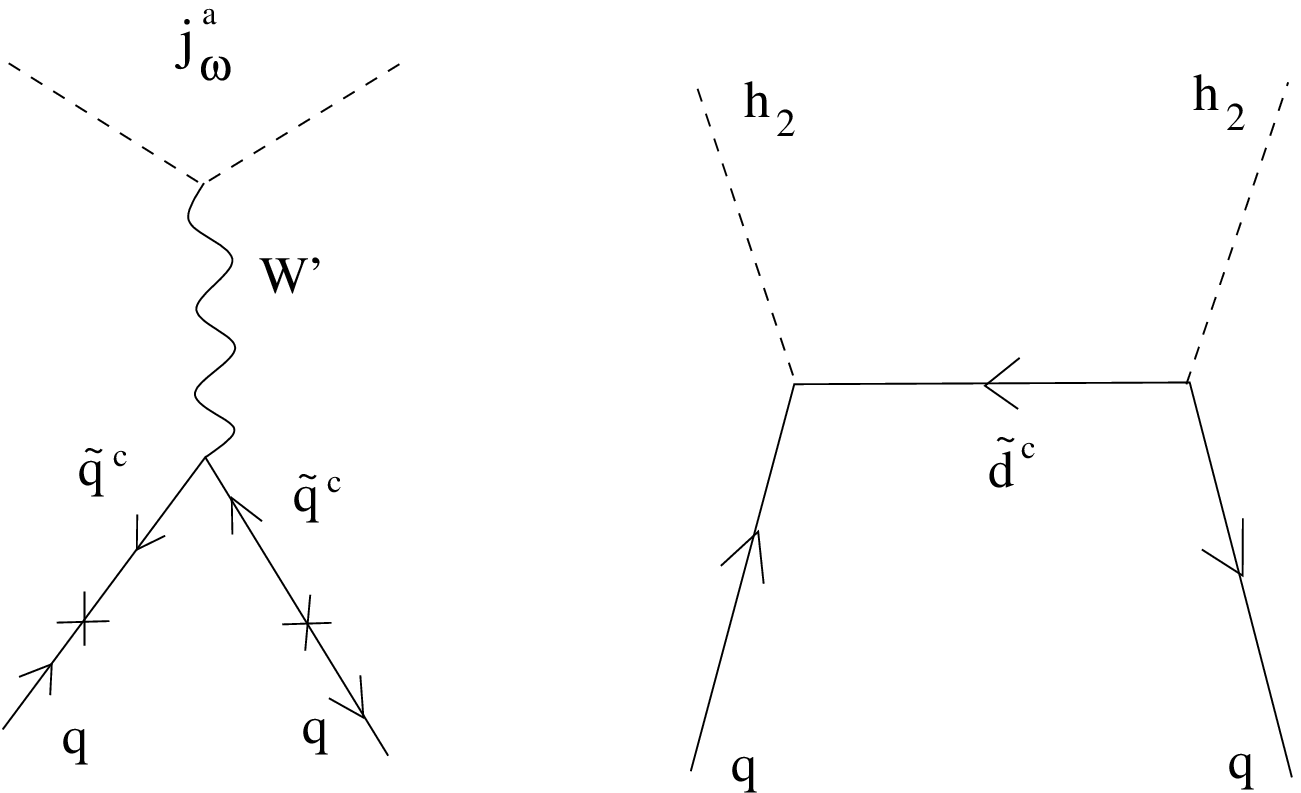, width=2.7 in}
\label{Fig: Q3Diagrams}
\caption{Dominant diagrams that contribute to $\OO_{H q_3} ^L$ and $\OO_{H q_3}^Y$
at tree level.}
\end{figure}

The following third-generation operators can 
be probed by precision tests:
\begin{eqnarray}
\OO_{H q_3}^L=j_{\omega\; a \mu} \bar{q}_3 \sigma^a \bar{\sigma}^\mu q_3 
\hspace{0.3in}
\OO_{H q_3}^Y=j_{\omega\; \mu} \bar{q}_3 \bar{\sigma}^\mu q_3 
\hspace{0.3in}
\OO_{H d^c_3}^Y= j_{\omega\; \mu} \bar{d}^c_3 \bar{\sigma}^\mu d^c_3. 
\label{eq:3ops}
\end{eqnarray}  
$\OO_{H q_3}^L$ receives an enhancement through the mixing of $q_3$ with
$\tilde{q}_1$ (see Fig.~\ref{Fig: Q3Diagrams}):  the $q_3$ coupling to the $W'$ is proportional to
$\tan \theta$, which was constrained to be relatively small, $\lsim 0.2$, 
from the precision electroweak considerations
of Sec. \ref{Sec: PEWCurrents}, but  $\tilde{q}_1$ is charged under the opposite
$SU(2)$ and couples as $\cot \theta \gsim 5$.  In terms of the doublet
mixing angle $\vartheta_Q$, the coefficient of this operator is
\begin{eqnarray}
c_{H q_3}^L = \frac{v^2 }{2 f^2}
\cos 2 \theta (\sin^2\theta \cos^2 \vartheta_Q
 - \sin^2\vartheta_Q \cos^2 \theta).
\end{eqnarray}
Other contributions  to the operators of Eq.~(\ref{eq:3ops}) come from the 
$\tilde{q}_1 h_2^\dagger \tilde{d}^c$ interaction of the top sector.  
Both naturalness and precision electroweak considerations
prefer somewhat small values of $\vartheta_Q$ for the top sector of 
Sec.~\ref{Sec: Top}, in which case the quark doublet is dominantly 
$\tilde{q}_1$ rather than $q_3$.  Thus this interaction produces relatively 
unsuppressed coefficients
for the  $\OO_{H q_3} ^L$ and $\OO_{H q_3}^Y$ operators, 
\begin{eqnarray}
c_{H q_3}^L = \frac{v^2}{f^2} \tan^2 \beta \cos^2 \vartheta_U
\hspace{0.5in}
c_{H q_3}^Y = \frac{3v^2}{f^2} \tan^2 \beta \cos^2 \vartheta_U.
\end{eqnarray}
The hypercharge operators receive corrections from the $B'$ exchange as well,
with coefficients
\begin{eqnarray}
\nonumber
&&c_{H q_3}^Y = \frac{v^2}{\bar{f}^2 } \cos 2 \theta' \sin^2\theta'
\simeq 2\delta \theta' \frac{v^2}{\bar{f}^2}\\
&&c_{H d^c_3}^Y =  \frac{v^2}{\bar{f}^2} \cos 2 \theta' \sin^2\theta'
\simeq 2\delta \theta' \frac{v^2}{\bar{f}^2},
\end{eqnarray}
where we take $\theta'=\frac{\pi}{4}+\delta \theta'$.
In Sec.~\ref{Sec: Top} additional mixing terms were briefly mentioned
that may induce addional effects that were ignored.  Finally, radiative 
contributions to these operators from the Higgs, the top and the $W'$ may also 
be important.  

Another interesting aspect of third-generation physics is the flavor
mixing mentioned earlier in Sec.~\ref{Sec: Top}.
While the FCNCs mediated by the $Z^0$ and $B'$ seem to be small
with appropriate choices of Yukawa couplings, this issue is beyond
the scope of this paper and needs to be considered in more depth.
%Another potentially interesting source of flavor violating operators are 
%those that consider $b \rightarrow s \gamma$ arising from various sources.   

\subsection{$B'$ Production}
\label{Sec: B' Production}

Equation \ref{Eq: B' Mass} shows that even in the presence of an
additional breaking scale $F$, the $B'$ can be quite light in the 
near-oblique limit, 
%$\theta' \rightarrow \frac{\pi}{4}$, 
with a mass of $375 \GeV$ 
for $\bar{f} = 2 \TeV$. Colliders have already begun to probe 
this energy scale and there is no evidence for a $B'$.  
Is it possible that colliders could have missed a vector of this mass?  
Precision electroweak constraints suggest that the coupling of the 
$B'$ to fermions is quite suppressed based in the present model,
in which case the $B'$ can evade detection.
Taking $\theta' = \frac{\pi}{4} +  \delta \theta'$ with 
$\delta \theta' \lsim \frac{1}{10}$, 
%to satisfy precision electroweak constraints, 
the coupling to the light fermions is
\begin{eqnarray}
\LL_{\text{Int}} \simeq 
- 2\, \delta \theta'\, g'\, B'_\mu\, j_{\text{F}}^\mu .
\end{eqnarray}
This means that the production rate from accelerators is very small
and the decay width into the first two generation is very small as well.
Because the third generation is charged only under a single $U(1)$,
it couples significantly to the $B'$ and completely dominates
the decay width.  Thus the Drell-Yan production of the $B'$ is
suppressed and the decay width into the electrons and muons is 
small.  It requires looking at the tau channel to see  
the $B'$ and the limits are less constraining.

As emphasized earlier, it is always possible to gauge just the 
diagonal $U(1)_Y$ rather than a product of $U(1)$'s, in which case
the constraints associated with the $B'$ are removed.  

\subsection{Summary of Results}
\label{Sec: Results}

\begin{figure}[ht]
\centering\epsfig{figure=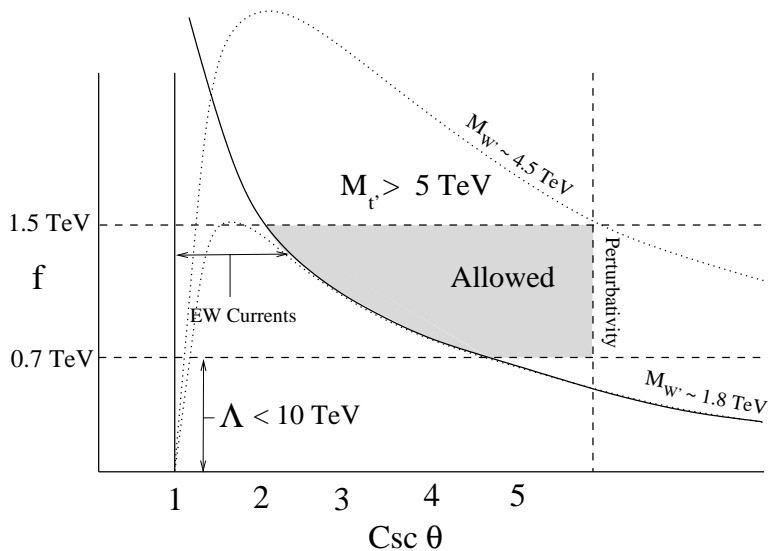, width=4.0 in}
\caption{
\label{Fig: Limits}
Sketch of the parameter space
that gives adequately small non-oblique corrections. The bounds
on $f$ come from naturalness considerations (upper) and the
requirement that the cutoff satisfies $\Lambda \gsim 10$~TeV (lower).
}
\end{figure}

In our discussion we have emphasized the usefulness of the near-oblique
limit for identifying regions of parameter space that lead to acceptably
small precision electroweak corrections.  How close to this limit
do we need to be?  Focusing on the $SU(2)$ interactions,
a rough answer to this question is given in Fig.~\ref{Fig: Limits}.  
From this figure one can infer, for a given value of $f$, the range
of values of the mixing angle $\theta$ that give small non-oblique
corrections.  We require $f$ to be above 700~GeV to keep the cutoff
near 10~TeV (and also because for smaller values of $f$ the constraints
on  $\theta$ and $\theta'$ become increasingly severe), while the 
upper bound on $f$ is motivated by naturalness considerations.
The plot also shows a lower bound on $\theta$ from the requirement that
$g_2$ remain perturbative.  The mass
of the $W'$ ranges from roughly 1.8 to 4.5~TeV in the allowed region.

\begin{figure}[ht]
\centering \epsfig{figure=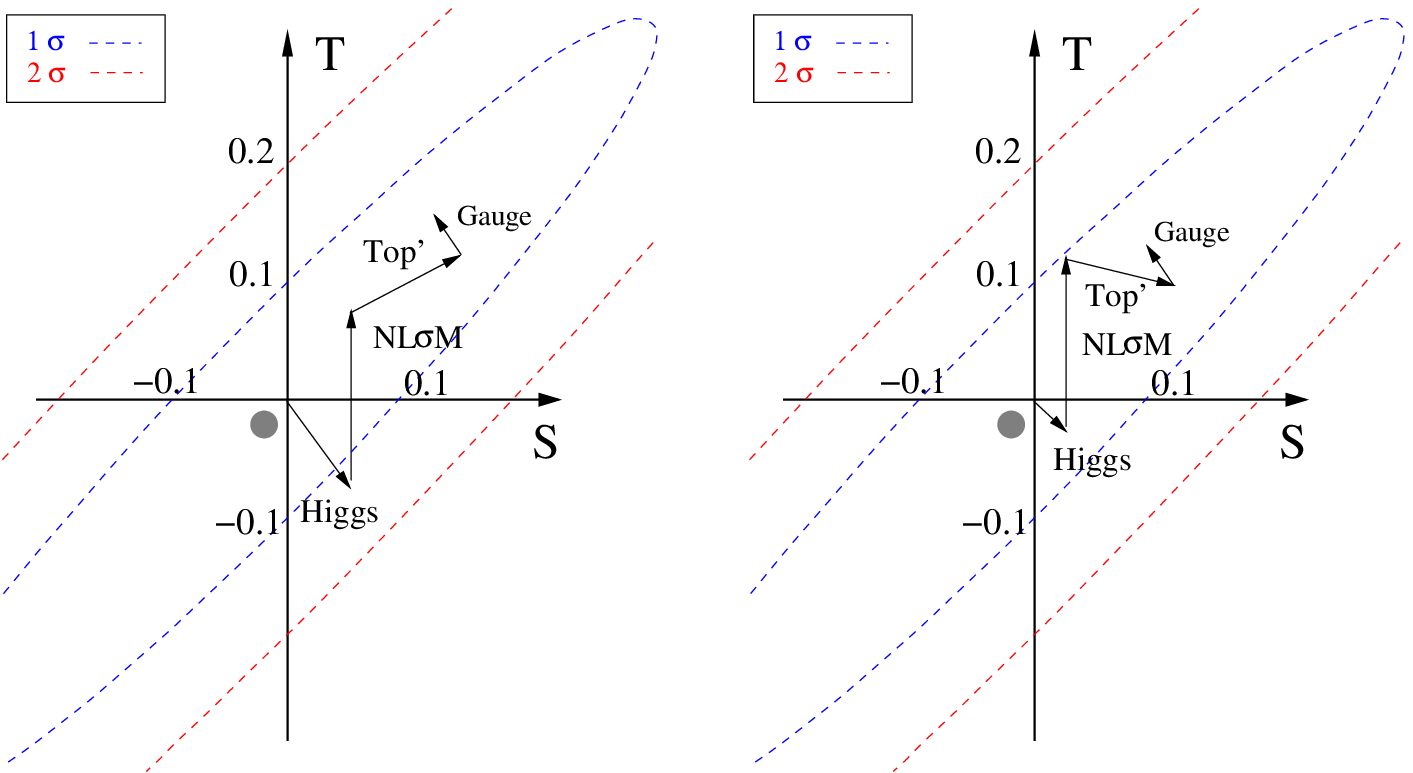, width=5.2in}
\caption{
\label{Fig: STLimits}
Contributions to the $S$ and $T$ parameters for the reference parameter values 
of Sec.~\ref{Sec: Overview}. The dashed ellipses are roughly the 
$1 \sigma$ and $2 \sigma$ limits in the $S-T$ plane.}
\end{figure}

For points inside the allowed region of this plot, non-oblique
corrections are sufficiently small that an analysis of the oblique corrections
using the $S$ and $T$ parameters is meaningful.  In Sec.~\ref{Sec: Oblique}, 
we calculated the contributions to $T$ from the gauge interactions (positive
and small in the near-oblique limit, or absent entirely if only $U(1)_Y$ is gauged), 
the non-linear sigma model self-interactions 
(positive and small for $\tan \beta \simeq 1$), Higgs loops 
(typically negative, and growing in magnitude with the the quartic coupling 
$\lambda$),
and top-sector loops (reasonably small for parameters that give mild 
radiative corrections to the Higgs mass squared, as preferred by naturalness 
considerations). We also calculated the contributions to $S$ and found sizable
positive contributions from the top and Higgs sectors, 
so a positive contribution to $T$ may in fact be welcome given 
the shape of the $S-T$ ellipse.

In Fig.~\ref{Fig: STLimits} we show the $S-T$ ellipse along with the various 
contributions to $S$ and $T$ that arise for both sets of reference values 
given in 
Sec.~\ref{Sec: Overview}.  The total contributions are in reasonable agreement 
with precision data.  Certainly there are different parameter choices that 
lead to much larger corrections, but we believe that it is an important 
result that there is no intrinsic conflict between the naturalness of the 
Higgs potential and precision electroweak data in this theory.  In the next 
section we give a more thorough consideration of this issue.
%%%%

%%%%
\section{Discussion}
\setcounter{equation}{0}
\renewcommand{\theequation}{\thesection.\arabic{equation}}
\label{Sec: Conc}

\subsection{Naturalness, Precision Electroweak, and Higgs Physics}

In this section we explore the interplay between precision electroweak
constraints, naturalness considerations, and Higgs physics in this
model.

We focus on values of $\tan \beta$ close to one, giving small oblique
corrections from the non-linear sigma model self interactions.
We have already argued that these values are quite natural in 
this model given the approximate $SU(2)_{H,R}$ symmetries.  One can express 
$\tan \beta$ in terms of $\delta m^2$, the calculable and finite radiative 
Higgs mass from the top sector, and $\Delta m^2$, the incalculable but 
$SU(2)_{H}$-symmetric contributions from the gauge and scalar sectors 
(Eq.~\ref{Eq: tanbeta}).  Choosing top-sector angles $\vartheta_Q$ and 
$\vartheta_U$ that minimize the top contribution, $\delta m^2$ can in turn be 
rewritten in terms of  $\beta$ and $f$ (Eq.~\ref{Eq: SU6 Rad Cor2}), and we 
can finally approximate Eq.~\ref{Eq: tanbeta} as
\begin{eqnarray}
\cos 2 \beta(1 + \cos 2 \beta) &\simeq& \frac{f^2}{8 \Delta m^2},
\label{Eq: cos2beta}
\end{eqnarray} 
where we have expanded around  $\cos 2\beta=0$.  Naturalness motivates us to 
concentrate on relatively small values for $\Delta m^2$, but if we take it too 
small, $\cos 2\beta$ becomes too large based on precision electroweak 
considerations.  For our reference value $f=700$~GeV, taking 
$\Delta m=550$~GeV   gives $\cos 2\beta \simeq 0.18$, and
a reasonably small non-linear sigma model contribution to $T$  according to the
discussion in Sec.~\ref{Sec: NLSM}.  The same parameters give $\delta m \simeq300$~GeV.

Before proceeding further, we note that this value for $\Delta m$ is in rough
agreement with what one would expect based on the logarithmically divergent 
one-loop gauge and scalar 
contributions.  For $f=700$~GeV, Fig.~\ref{Fig: Limits} shows that 
non-oblique corrections require $\theta\simeq 1/5$, in which case  
%Setting $\Lambda \simeq 4 \pi f$ in the logarithm,
Eq.~(\ref{Eq: gaugerad}) gives
\begin{eqnarray}
\Delta m^2_{\text{$SU(2)$ Gauge}} \simeq  (300 \GeV )^2, 
\end{eqnarray}
while the $U(1)$ gauge contribution is negligible even with a breaking 
scale $\bar{f} \sim 2$~TeV.  Finally, the scalar contribution of Eq.~(\ref{Eq: scalarrad})
is 
\begin{eqnarray}
\Delta m^2_{\text{Scalar}} \simeq \lambda^2 (200 \GeV)^2.
\end{eqnarray}
As always, naturalness prefers larger values of $\lambda$, and depending
on the contribution to $T$ from the top sector, having $\lambda \sim 2 - 4$
may be favorable for agreement with precision data.  However even for
smaller $\lambda$ the large cut-off uncertainties in these estimates
make it completely reasonable to have $\Delta m \gsim 500 \GeV$.

For the values of $\delta m$ and $\Delta m$ arrived at above, we 
find that the fine-tuning parameter 
introduced in Eq.~\ref{Eq: Intro FT} is
\begin{eqnarray}
\kappa \simeq 0.06 \,\lambda.
\end{eqnarray} 
For a positive contribution to $T$ from the top sector, as arises
for the first set of reference parameters in Sec. \ref{Sec: Overview},
large values  of $\lambda$ are welcome because of the compensating 
corrections from the Higgs sector.  In this case, the fine tuning
is quite mild and lightest Higgs boson is moderately heavy.  This
is not to say that a light Higgs boson is disfavored.   In fact,
some parameters of the top sector, such as those in the second set
of reference parameters of Sec. \ref{Sec: Overview}, give rise
to a negative contribution to $T$ and a sizeable oblique correction
from the Higgs sector is unnecessary.   Small $\lambda$ and a light
Higgs boson are completely consistent with precision data, although
the fine tuning becomes more severe as $\lambda$ is decreased. 

\subsubsection*{Positive $\Delta T$ from the Top Sector}

\begin{figure}[ht]
\centering
  \epsfig{figure=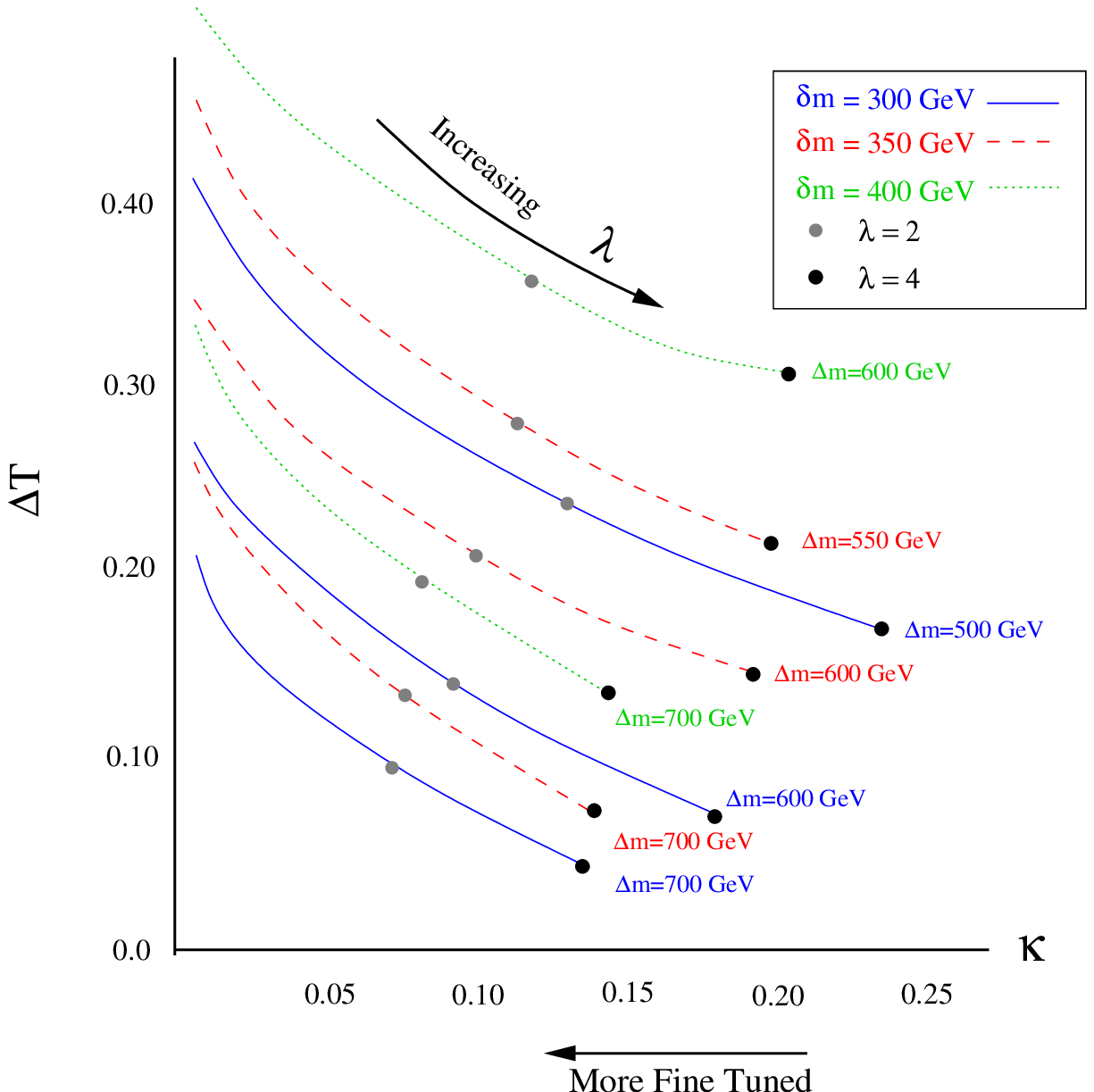,width=4.2in}
\caption{
\label{Fig: FTLambda}
$T$ vs. the  fine-tuning parameter $\kappa$, for several values of $\delta m$ 
and $\Delta m$, as the quartic coupling $\lambda$ is varied.
For a given value of $\delta m$,  naturalness prefers small
$\Delta m$ while precision electroweak data prefer larger values.
Increasing the quartic coupling significantly reduces both the fine tuning and
$\Delta T$.}
\end{figure}

\begin{figure}[ht]
\centering
  \epsfig{figure=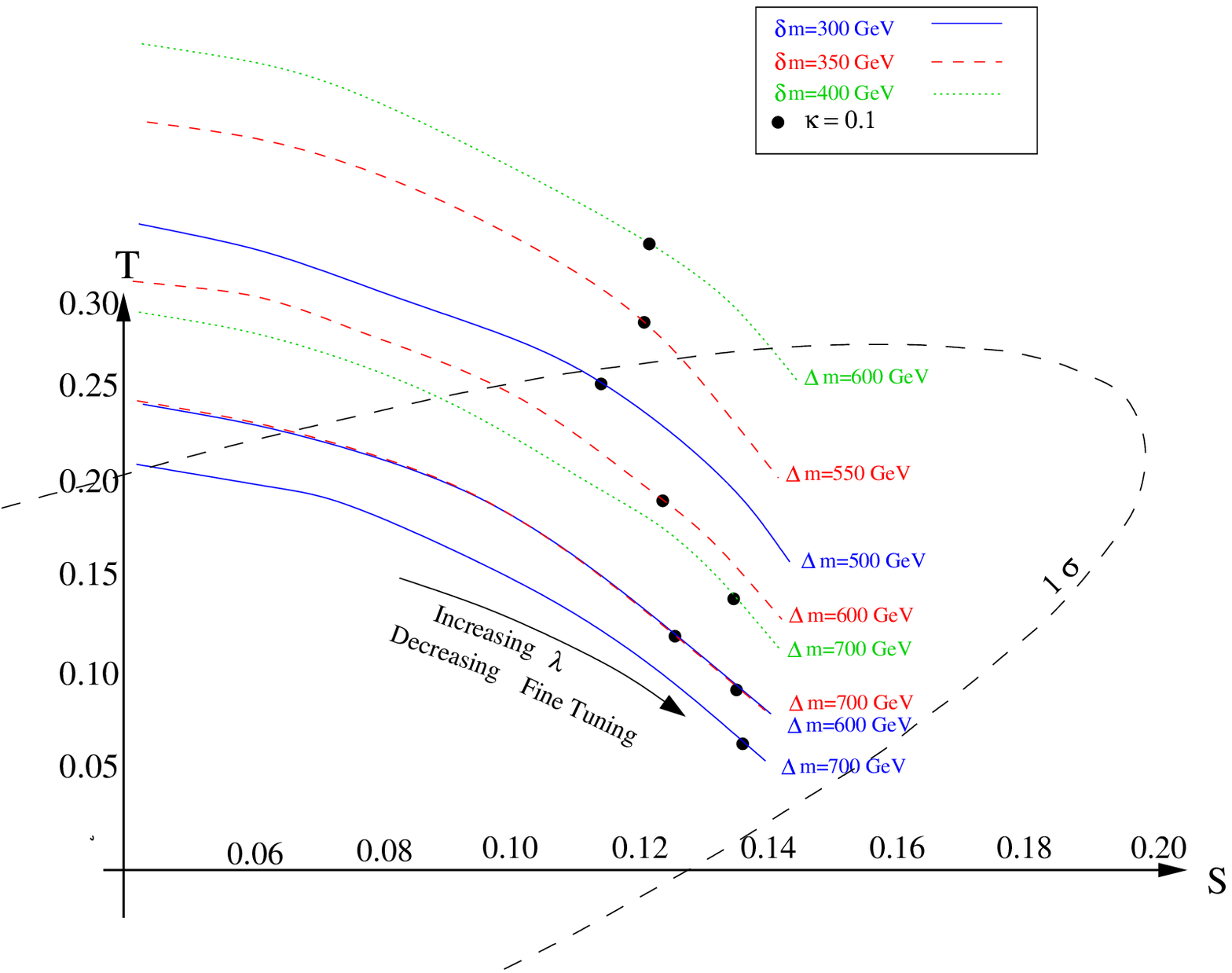,width=4.2in}
\caption{
\label{Fig: STFT}
$S-T$ contours for the same values of $\delta m$ and $\Delta m$ as in 
Fig.~\ref{Fig: FTLambda}.  The bullets denote 10\% fine tuning and the 
one-sigma ellipse is shown.}
\end{figure}

These trends are depicted in Figs.~\ref{Fig: FTLambda}  through 
\ref{Fig: STFT2} 
for various $\delta m$ and $\Delta m$.  In these plots the contributions to
$S$ and $T$ include all of those calculated in Sec.~\ref{Sec: PEW}.  
For the gauge contributions we take $F=2$~TeV, $\delta \theta'=0.1$ and
$\theta=0.2$ to be sufficiently close to the near-oblique limit.  Eliminating 
the $B'$ by gauging only $U(1)_Y$ shifts the contours downwards by 0.04 in $T$.

In Figs. \ref{Fig: FTLambda} and \ref{Fig: STFT},
the top contributions are calculated taking the first set of reference 
values $\cos^2\vartheta_U=\cos^2\vartheta_Q=2/3$.\footnote{
By keeping these angles fixed we get a slight mismatch between the values 
of $\delta m$ as given by the top sector and the values used as inputs, a 
small discrepancy we ignore because of the $\OO(20\%)$ uncertainty in 
$\delta m^2$ discussed in Sec.~\ref{Sec: EWSB}.}  
In Fig.~\ref{Fig: FTLambda} we see that for the Higgs mass parameters 
considered, $10-20$\% fine tuning is required for $\lambda\sim 2-4$.  
The net contributions to $T$ can be reasonably small, but are certainly 
far from negligible. As shown in Fig.~\ref{Fig: STFT}, increasing $\lambda$ 
also increases the Higgs sector's contribution to $S$, helpful given the 
positive contributions to $T$ and the shape of the $S-T$ ellipse.  The plot 
shows that there are indeed parameters that give  decent agreement
with precision data and only mild fine tuning.

For illustrative purposes, let us consider the Higgs spectrum for the 
parameters from above, $\Delta m=550$~GeV and $\delta m=300$~GeV, taking 
$\lambda=3$.  The fine tuning is given by $\kappa=0.18$ and the oblique 
corrections are $(S,T)=(0.13,0.13)$.  Using the fact that $\cos 2\beta$ 
is small, the mass of the lightest Higgs from Eq.~\ref{Eq: Higgs Masses} 
can be approximated as
\begin{eqnarray}
m^2_{h^0} = \frac{\lambda v^2}{2} (1 + \OO(\cos^2 2\beta)).
\end{eqnarray}
Numerically we find 
\begin{eqnarray}
m_{h^0} =  297 \GeV
\hspace{0.3in}
m_{H^0} = 720  \GeV
\hspace{0.3in}
m_{H^\pm} = 718 \GeV
\hspace{0.3in}
m_{A^0} =  779 \GeV .
\end{eqnarray}
The charged Higgs and heavier CP even Higgs are typically
nearly degenerate in the regions of parameter space that
are preferred by precision electroweak data, given that 
$\kappa \lsim \frac{1}{3}$.

From Eq.~\ref{Eq: Alpha} one can see that the $h^0 - H^0$ mixing angle 
$\alpha$ is typically close to $- \beta$ because $x \simeq 1 - \kappa$.
This means that the $h^0$ and $H^0$ couplings to fermions have no large 
enhancements or suppressions,  because both angles are $\OO(1)$.  The 
couplings of the $W^\pm$ and $Z^0$ to $h^0$ go like
$\sin(\beta -\alpha) \sim \sin 2 \beta$ and are essentially unsuppressed,
while the couplings to $H^0$ are proportional to  
$\cos (\beta - \alpha) \sim \cos 2\beta$, and are quite suppressed. 
Thus the lightest Higgs looks very much like the Standard Model 
Higgs in terms of its couplings.

\subsubsection*{Negative $\Delta T$ from the Top Sector}

\begin{figure}[ht]
\centering
  \epsfig{figure=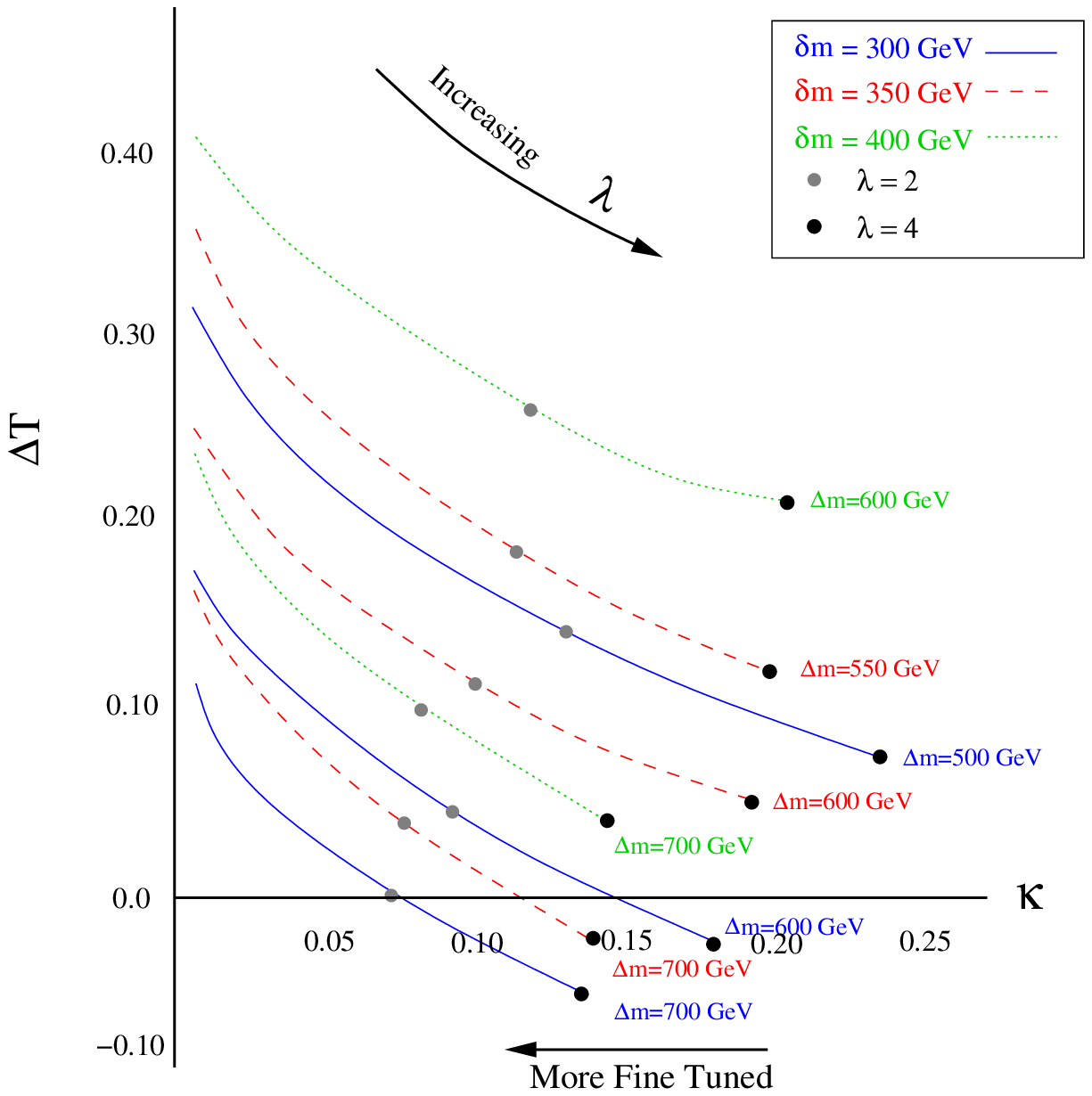,width=4.2in}
\caption{
\label{Fig: FTLambda2}
$T$ vs. the  fine-tuning parameter $\kappa$, for several
values of $\delta m$ and $\Delta m$, as the quartic coupling $\lambda$ 
is varied.
For a given value of $\delta m$,  naturalness prefers small
$\Delta m$ while precision electroweak data prefer larger values.
Increasing the quartic coupling significantly reduces both the fine tuning 
and $\Delta T$.}
\end{figure}

\begin{figure}[ht]
\centering
  \epsfig{figure=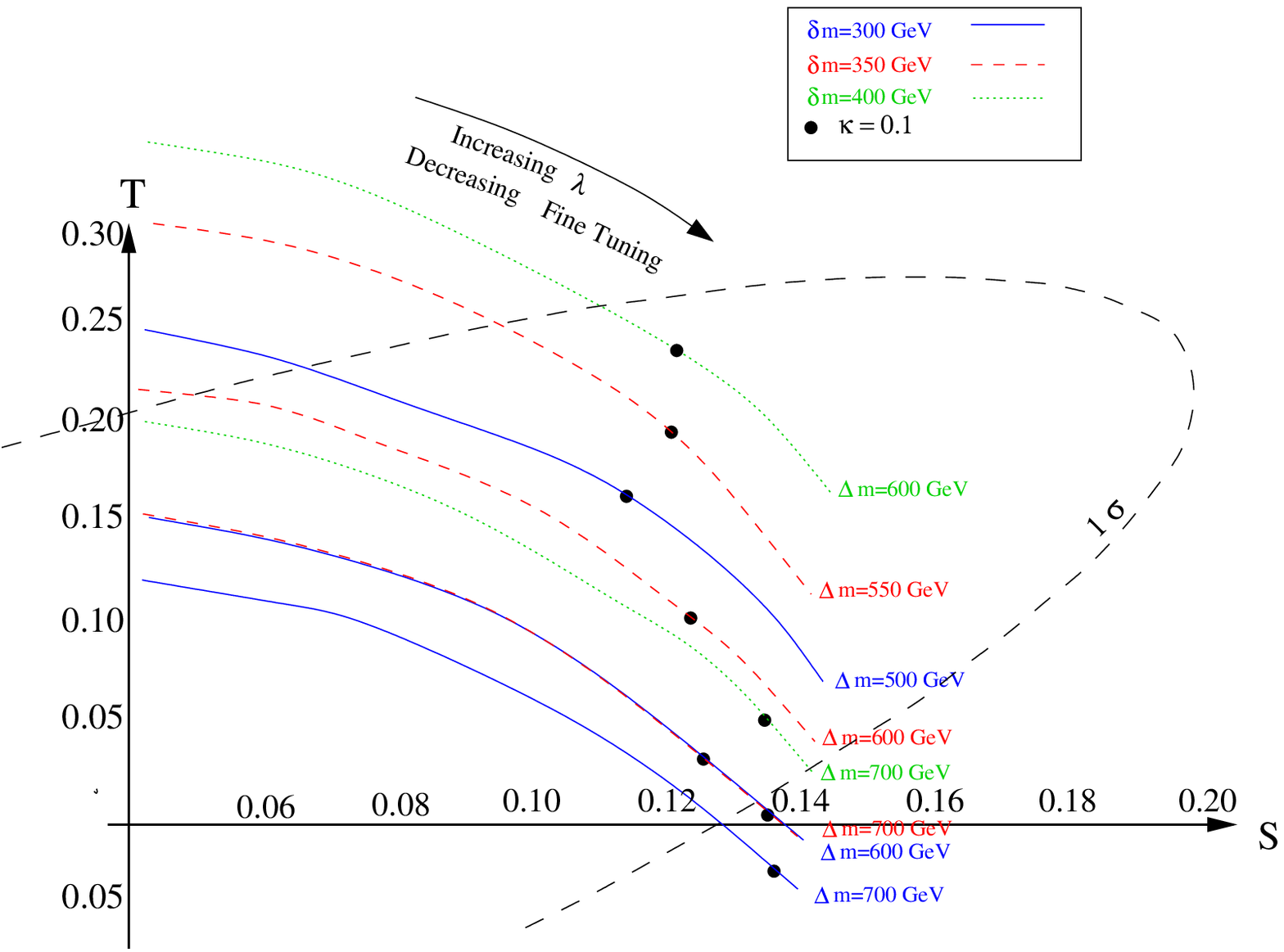,width=4.2in}
\caption{
\label{Fig: STFT2}
$S-T$ with a negative top contribution to $T$.  The contours are for the 
same values of $\delta m$ and $\Delta m$ as in Fig.~\ref{Fig: FTLambda2}.  
The bullets denote 10\% fine tuning and the one-sigma ellipse is shown.}
\end{figure}

In Figs.~\ref{Fig: FTLambda2} and \ref{Fig: STFT2} 
we use the second set of reference values for the top sector, 
$\cos^2\vartheta_U=3/4$ and $\cos^2\vartheta_Q=3/5$.
The top sector gives a negative contribution to $T$ for these
values, and in this case there is no tension between 
small $\lambda$ and precision data.

Taking the $\Delta m^2$ and $\delta m^2$ as before, but now with 
$\lambda = 0.5$, we find 
\begin{eqnarray}
m_{h^0} =  122 \GeV
\hspace{0.3in}
m_{H^0} = 718  \GeV
\hspace{0.3in}
m_{H^\pm} = 718 \GeV
\hspace{0.3in}
m_{A^0} =  728 \GeV .
\end{eqnarray}
Notice that heavy Higgs spectrum is insensitive to $\lambda$.
As before, the couplings of the lightest Higgs boson are Standard Model-like.

\subsection{Conclusions and Outlook}

In this paper we studied precision electroweak constraints on 
the $SU(6)/Sp(6)$ little Higgs theory.  We found it useful
to first identify a ``near-oblique limit'' in which the heavy
$W'$ and $B'$ of the model decouple from the light two generations
of fermions.  Then we calculated oblique corrections that arise from
the non-linear sigma model self-interactions, gauge interactions,
top loops, and Higgs loops.  We found parameter space with
$(S,T)$ in reasonable agreement with precision data, and with only
mild fine tuning in the Higgs sector, at the $10-20$\% level.
Non-oblique corrections involving the third generation were considered
only briefly and deserve further study.

In our study we explored a number of possible modifications
to the model as presented in \cite{Low:2002ws}.  First,
we noted that an independent breaking scale for the $U(1)$ gauge
interactions could be introduced without affecting other aspects
of the model -- in fact, it is possible to remove the $B'$ from the
theory altogether, by gauging only the Standard Model hypercharge,
without introducing severe fine tuning.  The additional
breaking may be necessary for values of $f$ consistent with
naturalness, in order to evade direct constraints from $B'$
production.  A second modification was to charge the light two generations equally 
under both $U(1)$ groups  in order to realize the near-oblique limit.
Since the third generation is charged only one $U(1)$ this
leads to flavor non-universality in the couplings of the $B'$, and
the possibility of interesting flavor-changing signatures.  
Finally, we considered a new top sector that gives finite
radiative contributions to the Higgs potential.  In the appendix
we also consider other alternatives for the top sector, and the
oblique corrections induced by the third generation fermions change
depending on the setup.

Tree-level oblique corrections arising from the non-linear sigma model
structure of the theory vanish for $\tan \beta=1$. In the absence
of radiative corrections from the top sector, approximate $SU(2)$
symmetries of the model guarantee that $\tan \beta$ is very close to one.
%The top contributions take $\tan \beta$ away from one.  
In the top sector of Sec.~\ref{Sec: Top}, values for the mixing angles
$\vartheta_Q$ and $\vartheta_U$ that minimize the top contribution to the 
Higgs mass, $\delta m^2$,  tend to give small contributions to
the $T$ parameter. These smaller values of $\delta m^2$ reduce
the minimum gauge and scalar radiative contributions 
necessary for $\tan \beta$ to be consistent with precision
electroweak constraints, which in turn reduces the fine tuning
in the Higgs potential. For $f$ ranging from $0.7-1.5$~TeV, 
the values of the masses that minimize $\delta m^2$ are  
$m_{u_H}, m_{q_H}\simeq 2-4$~TeV, while the other particles
of the top sector have masses $m_{d_H}, m_{q'_H}\simeq 1-2$~TeV for the
same parameters.  

For the same range of $f$, the mass of the $W'$ lies within $\simeq
1.8-4.5$~TeV for parameters that give adequately small non-oblique 
corrections. In the presence of the extra breaking
scale, the mass of the $B'$ is not tied to the others, 
but even for ${\overline f}$ as large as 2~TeV, it is quite light,
with a mass of only 375~GeV.  To adequately suppress non-oblique corrections
associated with a $B'$ of this mass it must couple only weakly to light 
fermions, which complicates collider searches.

We found that for some top sector parameters, 
oblique corrections from Higgs loops improve
the agreement with precision data for somewhat large values of
the quartic coupling $\lambda$.  Larger $\lambda$ corresponds
to less severe fine tuning, so in this case 
precision data and naturalness considerations have similar
preferences.  In this case the lightest Higgs boson will be
somewhat heavy, though still less that $350 \GeV$ given
that we expect $\lambda \le 4$.
On the other hand, for other top sector parameters that give negative
contributions to $T$, small $\lambda$ is equally acceptable for
precision data.    In this case the mass of the lightest Higgs can
be near its current experimental bound.   In general, the 
tree-level couplings of the
lightest Higgs resemble those of the Standard Model Higgs.  The other
Higgs particles have masses of roughly 700 GeV or heavier irrespective
of the quartic coupling, with the pseudoscalar being the heaviest state.

%%%%

\section*{Acknowledgments}

We would like to thank N. Arkani-Hamed, S. Chang, and C. Csaki for 
many useful discussions during the course of this work.  
J.G.W.  would like to thank T. Rizzo for discussion on the physics
of the $B'$ in this model.
We would
like to thank R. Mahbubani and M. Schmaltz for reading an early draft 
and providing useful feedback.

The work of D.R.S. was supported by the U.S. Department of Energy under 
grant DE-FC02-94ER40818.

%%%
\appendix
\section{Alternate Top Sectors}
\setcounter{equation}{0}
\renewcommand{\theequation}{\thesection.\arabic{equation}}
\label{App: Other Tops}

In this appendix we consider alternatives to the top sector of 
Sec.~\ref{Sec: Top}, which have different radiative properies and
give different oblique corrections.
In \cite{Low:2002ws} a top sector with two separate couplings
to the non-linear sigma model field was introduced,
\begin{eqnarray}
\nonumber
\LL_{\text{top LSS}} &=&
y_1 f 
\left(\begin{array}{cccc}\tilde{q}& q_3&\tilde{u}&0\end{array}\right) 
\Sigma 
\left(\begin{array}{c}0\\0\\0\\u^c_3\end{array}\right) 
+y_2 f
\left(\begin{array}{cccc}0&q_3&0&0\end{array}\right)
\Sigma^*
\left(\begin{array}{c}
\tilde{q}^c\\0\\\tilde{u}^c\\\tilde{d}^c
\end{array}\right)\\
&&+
y_3 f \tilde{q}^c \tilde{q} + y_4 f \tilde{u}^c \tilde{u} +
y_5 f \tilde{d}^c \tilde{d}.
\label{eq:ts1}
\end{eqnarray}
The motivation for including both couplings was that it broke the Peccei-Quinn 
symmetry in the Higgs sector.  But as discussed in \cite{Low:2002ws}, if the 
Peccei-Quinn symmetry is broken elsewhere, simpler setups for the third 
generation are possible.    These are worth considering because the top 
sector of Eq.~(\ref{eq:ts1}) has five parameters, one combination
of which fixes the top mass while another fixes the $b$-term,
and a detailed analysis of this setup becomes quite complicated.  
To simplify the analysis, note that it is possible to decouple
either of the two interactions while keeping the top mass fixed
and leaving the Higgs sector radiatively stable.  The limits are
\begin{eqnarray}
y_2 \rightarrow 0 \hspace{0.5in} y_5 \rightarrow 0 \hspace{0.5in} \text{ $SU(5)_
L$ Minimal Top Sector}\\
y_1 \rightarrow 0 \hspace{0.5in} y_4 \rightarrow 0 \hspace{0.5in} \text{ $SU(4)_
R$ Minimal Top Sector}
\end{eqnarray} 
In the limit for the $SU(4)_R$ MTS, it is also necessary to decouple the 
additional light $SU(2)_L$ singlet with an interaction 
$\tilde{y}_4 f u_3^c\tilde{u}$. 
In this appendix, we study both of these top sectors.

\subsection{Summary of Other Top Sectors}

Before delving into the details of these different top sectors,
there are some general features that can be made.  The
structure of the radiative corrections to the Higgs mass is 
different than those of the full six-plet top sector that was
studied in the paper.  First, the contribution is log divergent
and there are two loop quadratic divergences that make $\delta m^2$,
the $SU(2)_H$ violating mass a parameter rather than a calculable
coefficient.   The structure of radiative corrections is of
the form
\begin{eqnarray}
V_{\text{1 loop top}} = 
- \frac{3 y_\text{top}^2}{8 \pi^2} m^2_{t'} \log\frac{\Lambda^2}{m^2_{t'}}
\left( |h_1|^2 + |h_2|^2\right)
+  \frac{3 y_\text{top}^2}{8\pi^2} \frac{m^2_{\tilde{t}'}}{\cos^2 \vartheta}\log\frac{\Lambda^2}{m^2_{\tilde{t}'}} |h_2|^2
\end{eqnarray}
where 
\begin{eqnarray}
m_{t'} \simeq \frac{2 y_\text{top} f}{\sin 2 \vartheta}
\end{eqnarray}
is the mass of the top partner canceling the quadratic divergence
from the  top quark.  Notice that the contribution proportional to this mass  is $SU(2)_H$ symmetric and
does not cause $\tan \beta$ to deviate from unity.  The second contribution that is not $SU(2)_H$ symmetric could in principle be made small by taking $m^2_{\tilde{t}'}$ small, making $\tan \beta$ very close to unity. This is possible because $m^2_{\tilde{t}'}$ is unrelated to the top Yukawa, as the $\tilde{t}'$ is responsible for canceling an auxiliary quadratic divergence that is not directly due to the Standard Model top quark loop, but to an interaction of the form 
$h_2^\dagger h_2 t^c \tilde{t}$ that appears when the top Yukawa coupling is covariantize. Unfortunately,  typically oblique corrections will
not allow the mass of this auxiliary top partner to be  
much smaller than $f$.

The next general feature is that the $SU(4)_R$ minimal top
sector typically gives a negative contribution to $T$ while
the $SU(5)_L$ minimal top sector gives a positive contribution.
Again there are log divergences to the $T$ parameter in the $SU(4)_R$ minimal top sector  and these
arise from the renormalization of the kinetic terms.  However
since the top sector does not preserve an $SU(6)$ chiral
symmetry, these log divergences typically renormalize non-$SU(6)$
invariant kinetic terms like $\Tr |\mathcal{P} D_\mu \Sigma|^2$
where $\mathcal{P}$ is a projection matrix.  These effects
could be important in naturalness considerations.  Throughout
the section the calculations have the log divergence cut-off
at $\Lambda \sim 4\pi f$. The $SU(5)_L$ minimal top sector, because it contains only heavy singlets, does not have log divergent contributions to the T parameter.

\begin{figure}[ht]
\centering
  \epsfig{figure=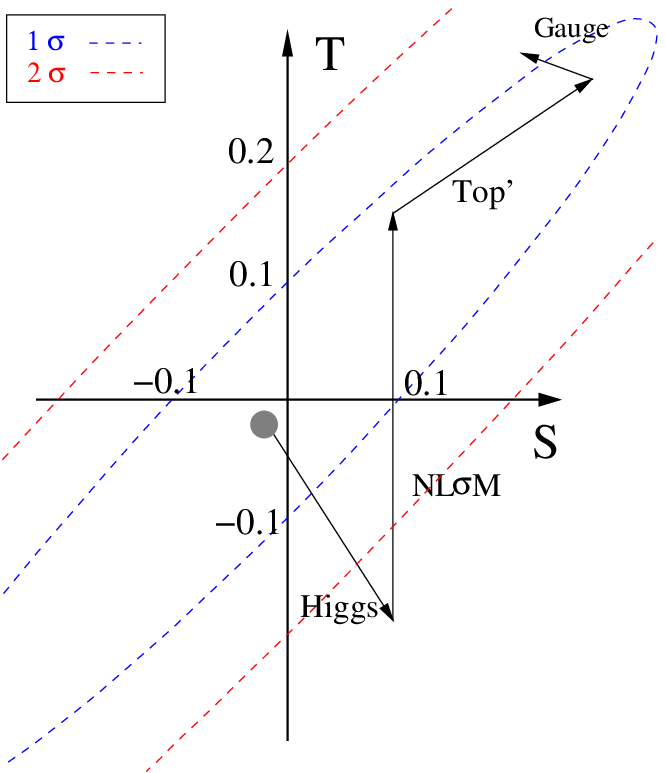,width=1.8in}
   \epsfig{figure=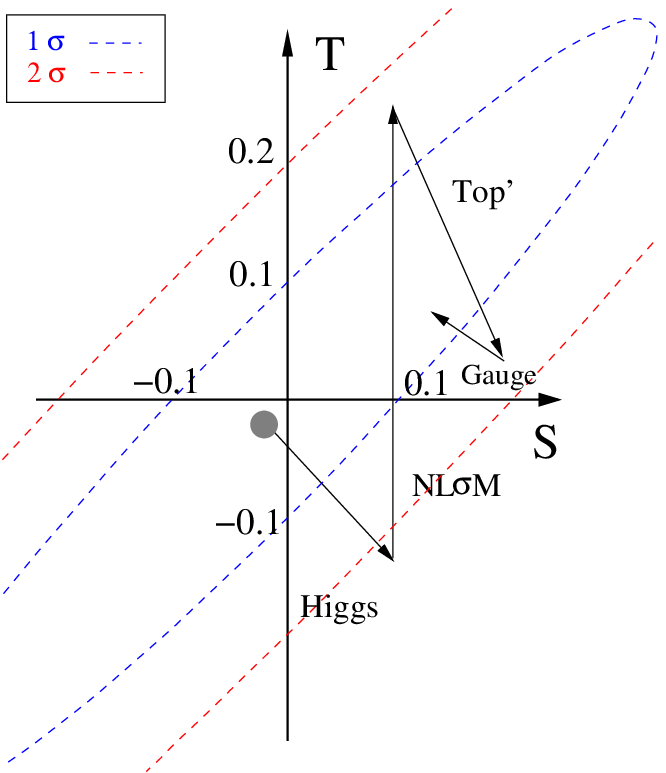, width=1.8in}
\caption{
\label{Fig: STLimits Alternate Top}
The dashed ellipses are roughly the $1 \sigma$ and $2 \sigma$ limits on 
the $S-T$ plane.
In the $SU(5)_L$ minimal top sector the contribution to $T$ is positive,
while in the $SU(4)_R$ minimal top sector the contribution to $T$ is negative.
The Higgs and top contributions are important for understanding
constraints on the model in these cases.
}
\end{figure}

\subsection{$SU(4)_R$ Minimal Top Sector}

The $SU(4)_R$ MTS adds a vector like quark doublet and down-like
singlet.  They combine with the third generation quark doublet and
up-like singlet to interact in an $SU(4)_R$ symmetric fashion.
This global symmetry prevents one loop quadratic divergences
from appearing from the top Yukawa coupling.
The fermions couple to $\Sigma$ as
\begin{eqnarray}
\LL_{\text{top}} =
y_1 f 
\left( \begin{array}{cccc}\tilde{q}^c&0&u_3^c&\tilde{d}^c\end{array} \right)
\Sigma^*
\left( \begin{array}{c}0\\q_3\\0\\0\end{array} \right)
+y_2 f\; \tilde{q}\tilde{q}^c 
+ \tilde{y}_2 f\; \tilde{d} \tilde{d}^c 
+\hc .
\end{eqnarray}
The quark doublet mixes with the 
vector-like fermion and can be diagonalized with mixing angles
$\vartheta_Q$,
\begin{eqnarray}
\tan \vartheta_Q = \frac{y_1}{y_2}
\hspace{0.5in}
y_{\text{top}}^{-2} = 
 |y_1|^{-2} + |y_2|^{-2},
\end{eqnarray}
while the masses are
\begin{eqnarray} 
m_{q_H} = \frac{ 2 y_{\text{top}} f }{\sin 2\vartheta_Q }
\hspace{0.5in}
m_{d_H} = \tilde{y}_2 f .
\end{eqnarray}
This top sector has the property that after electroweak symmetry breaking
the $\tilde{d}^c$ mixes with $d_3\subset q_3$.

\subsubsection{MTS Radiative Corrections}

The top Yukawa preserves an $SU(4)_R$ chiral symmetry, 
preventing a quadratically divergent mass term for the Higgs from being 
generated at one loop.
The one loop contribution from the Coleman-Weinberg potential is 
\begin{eqnarray}
\label{Eq: SU4R Rad Cor}
V_\eff =
 \frac{3 y^2_{\text{top}}}{4\pi^2}\left(
-m_{q_H}^2 \log \frac{\Lambda^2}{m_{q_H}^2} \; (|h_1|^2 +|h_2|^2)  
+\frac{ m_{d_H}^2}{\cos^2 \vartheta_Q} \log \frac{\Lambda^2}{m_{d_H}^2} \; |h_2|
^2  \right) .
\end{eqnarray}
This is the only interaction that breaks the $SU(2)_H$ symmetry
that rotates one Higgs doublet into the other.  As $m_d \rightarrow 0$
this contribution becomes smaller and the symmetry is restored.

\subsubsection{Oblique Corrections}

The minimal $SU(4)_R$ top sector has one heavy vector-like quark doublet and one
heavy vector-like down type quark singlet. The Standard Model top doublet
is a mixture of $q_3$ and $\tilde{q}$ with mixing angle $\vartheta_Q$. The
mass of the heavy single $d_H$ is given at first order by $\tilde{y_2}
f$. As mentioned above, the radiative corrections to the Higgs mass for this top sector have two
parts (see eq \ref{Eq: SU4R Rad Cor}). One is a common contribution to $m_1$
and $m_2$, proportional to $m_{q_H}$, which we denote $\Delta_{\text{top}} m=\sqrt{\Delta_{\text{top}} m^2}$ and the
other is a contribution to $m_1$ proportional to $m_{d_H}$, denoted $\delta
m=\sqrt{\delta m^2}$. $\delta m$ is positive, so it tends to create a $\tan \beta$ that is
larger than one.

\begin{figure}[t]
  \begin{center}
    \epsfig{figure=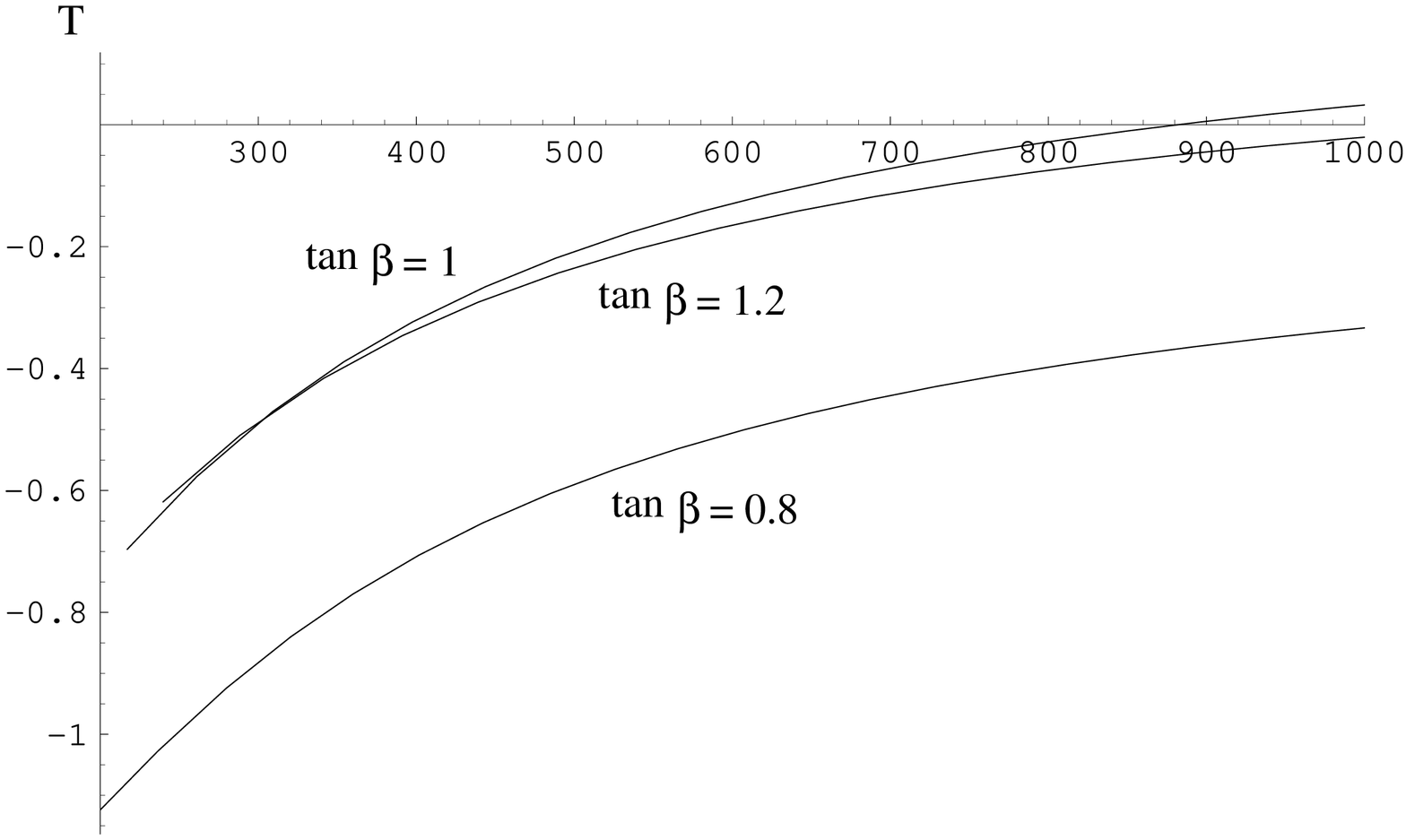,width=7cm} \epsfig{figure=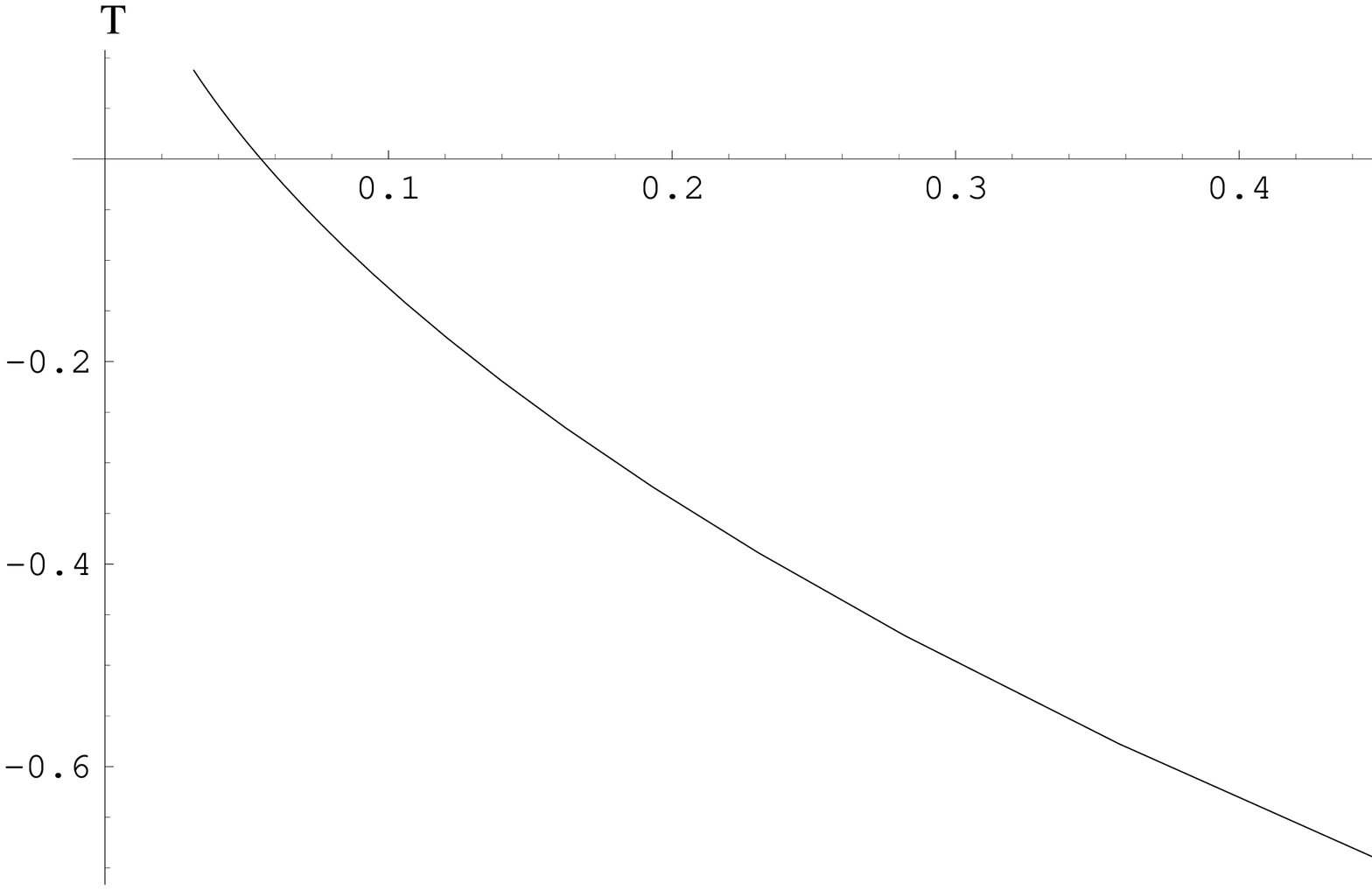,width=7cm}
    \caption{T as a function of $\delta m$ for $f=700 \GeV$, $\vartheta_Q =
  \pi/4$ for $\tan \beta$ of $0.8, 1$ and $1.2$. This correspond to $\Delta_{\text{top}}
  m$ of $506,617$ and $753$ $\GeV$ respectively. On the right hand side we
  show T vs S for $f=700 \GeV$, $\tan \beta=1$ and $\vartheta_Q = \pi/4$ }
  \label{Fig: T_MT4}  
  \end{center}
\end{figure}

 We find that the contribution to $T$ is in general negative. It
becomes more negative as $m_{d_H}$ is lowered, and do not depend very
strongly on $\vartheta_Q$. In figure \ref{Fig: T_MT4} we show T as a
function of $\delta m$ for $f=700 \GeV$, and
$\vartheta_Q = \pi/4$ assuming that the log in eq. \ref{Eq: SU4R Rad Cor}
is 5.  This is only an illustration of the typical
size of $\delta m$ as log divergent contributions are not calculable.
The choice of $\vartheta_Q$ minimize $\Delta m$ as
well as $\delta m$. We show also in figure \ref{Fig: T_MT4} $T$ vs $S$ for the $
f=700$, $\tan \beta=1$
and $\vartheta_Q=\pi/4$ when $\delta m$ is scanned over. We see that the
contributions to S can be quite important.

\subsection{$SU(5)_L$ Minimal Top Sector}

The $SU(5)_L$ MTS  adds an up-like singlet and a doublet to the
top sector.  
These couple to $\Sigma$ with the top quark, $q_3$ and $u_3^c$, through
\begin{eqnarray}
\LL_{\text{top}} =
y_1 f 
\left( \begin{array}{cccc}0&0&0&u_3^c\end{array} \right)
\Sigma
\left( \begin{array}{c}q\\\tilde{q}\\\tilde{u}\\0\end{array} \right)
+\tilde{y}_2 f\; \tilde{q}\tilde{q}^c 
+y_2 f\; \tilde{u} \tilde{u}^c 
+\hc
\end{eqnarray}
The quark doublet mixes with the 
vector-like fermion and can be diagonalized with mixing angles
$\vartheta_Q$,
\begin{eqnarray}
\tan \vartheta_U = \frac{y_1}{y_2}
\hspace{0.5in}
y_{\text{top}}^{-2} = 
 |y_1|^{-2} + |y_2|^{-2},
\end{eqnarray}
while the masses are
\begin{eqnarray} 
m_{q_H} = \frac{ 2 y_{\text{top}} f }{\sin 2\vartheta_Q }
\hspace{0.5in}
m_{u_H} =  \tilde{y}_2 f.
\end{eqnarray}

\subsubsection{Radiative Corrections}

The top Yukawa preserves an $SU(5)_L$ chiral symmetry, 
preventing a quadratically divergent mass term for the Higgs 
doublets from being
generated at one loop.
The one loop contribution from the Coleman-Weinberg potential is
\begin{eqnarray}
\label{Eq: SU5L Rad Cor}
V_\eff =
 \frac{3 y^2_{\text{top}}}{4\pi^2}\left(
-m_{q_H}^2 \log \frac{\Lambda^2}{m_{q_H}^2} \; (|h_1|^2 +|h_2|^2)  
+ \frac{m_{u_H}^2}{\cos^2 \vartheta_U} \log \frac{\Lambda^2}{m_{d_H}^2} \; |h_2|
^2  \right).
\end{eqnarray}
Only the second term breaks the $SU(2)_H$ symmetry
that rotates one Higgs doublet into the other.  As $m_u \rightarrow 0$
this contribution becomes smaller and the symmetry is restored.

\subsubsection{Oblique Corrections}

The $SU(5)_L$ minimal top sector has one heavy doublet, and one heavy up-type
singlet. At $0^{\text{th}}$ order, the doublet doesn't mix, and the
right-handed top is a mixture of $u_3$ and $\tilde{u}$ with mixing angle
$\vartheta_U$. Similarly to the $SU(4)_R$ minimal top sector, the radiative
corrections from the top sector gives a contribution common to $h_1$ and
$h_2$, $\Delta m$ proportional to $m_{u_H}$, and a contribution to $h_1$,
$\delta m$ proportional to $m_{q_H}$.

\begin{figure}[t]
\begin{center}
\epsfig{figure=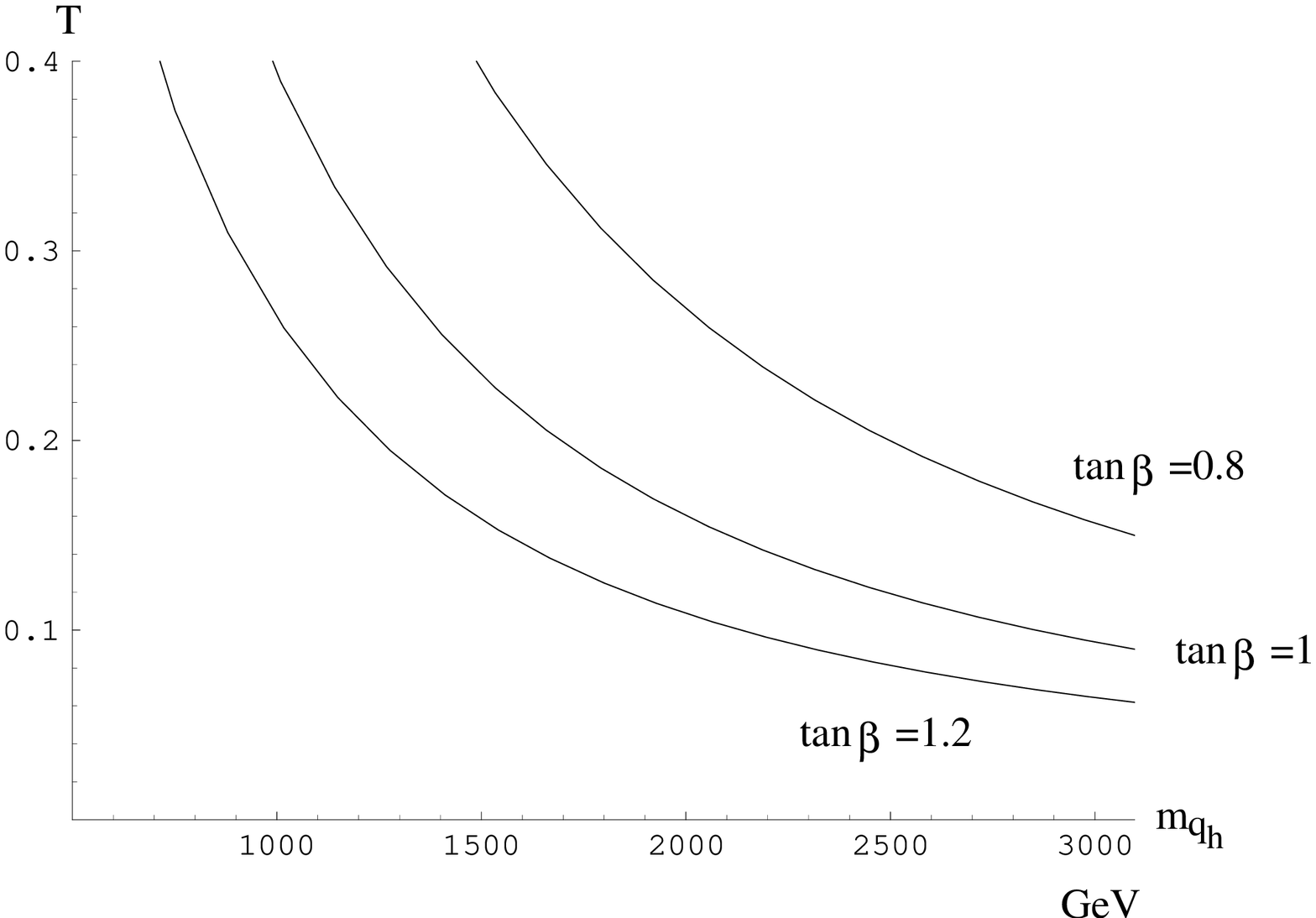,width=7cm} \, \epsfig{figure=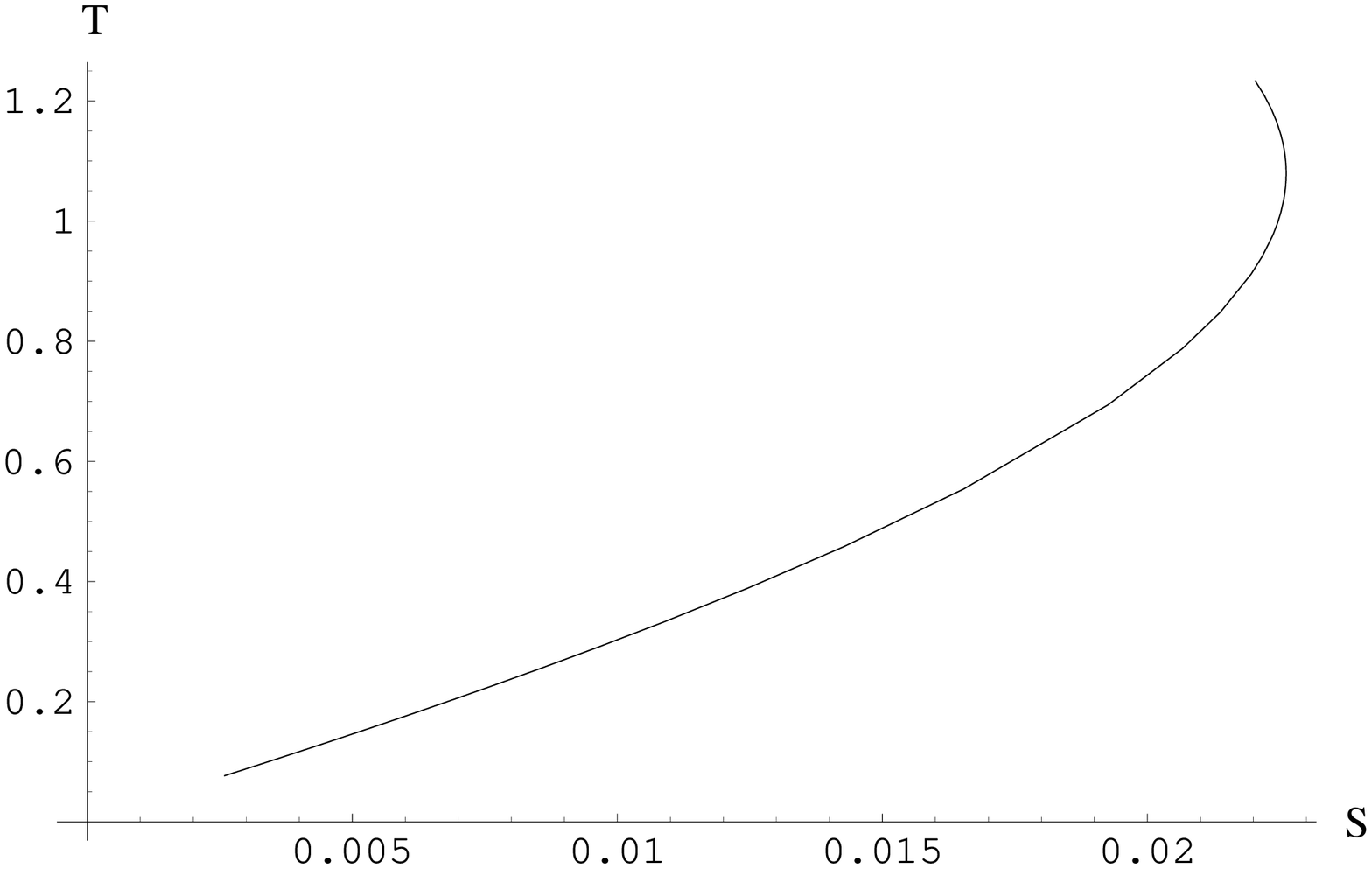,width=7cm}
\caption{
$T$ as a function of $m_{q_H}$ for $f=700 \GeV$, $m_{u_H}=2
\TeV$ and $\tan \beta$ of $0.8$, $1$ and $1.2$. On the right hand
side we show $T$ as a function of $S$ for $f=700 \GeV$, $\tan
\beta=1$, $m_{u_H} = 2 \TeV$ and $m_{q_H}$ going from $500 \GeV$ to
$3.5 \TeV$.   
\label{Fig: T_MT5}
}
\end{center}
\end{figure}

We find that the contribution to T in the $SU(5)_L$ minimal top sector are
quite important. They tend to decrease as $\vartheta_U$ is made small or as 
$y_2$
become large.  Both of these limits however, tend to increase the radiative
corrections to the Higgs mass. We show in figure \ref{Fig: T_MT5}, T as a
function of $m_{q_H}$ for $f=700 \GeV$ and various values of $\tan
\beta$. The dependence on $\vartheta_U$ is very mild.

We have also calculated the correction to the $S$ parameter and found it to
be quite small. Figure \ref{Fig: T_MT5} also shows $T$ as a function of $S$
for $f=700 \GeV$, $\tan \beta=1$, $m_{u_H} = 2 \TeV$ and a range of values
of $m_{q_H}$. 

%%%


\begin{thebibliography}{9}

%\cite{Arkani-Hamed:2001nc}
\bibitem{Arkani-Hamed:2001nc}
N.~Arkani-Hamed, A.~G.~Cohen and H.~Georgi,
%``Electroweak symmetry breaking from dimensional deconstruction,''
Phys.\ Lett.\ B {\bf 513}, 232 (2001)
%%[arXiv:hep-ph/0105239].
%%CITATION = HEP-PH 0105239;%%

%\cite{Arkani-Hamed:2002pa}
\bibitem{Arkani-Hamed:2002pa}
N.~Arkani-Hamed, A.~G.~Cohen, T.~Gregoire and J.~G.~Wacker,
%``Phenomenology of electroweak symmetry breaking from theory space,''
arXiv:hep-ph/0202089.
%%CITATION = HEP-PH 0202089;%%

%\cite{Arkani-Hamed:2002qx}
\bibitem{Arkani-Hamed:2002qx}
N.~Arkani-Hamed, A.~G.~Cohen, E.~Katz, A.~E.~Nelson, T.~Gregoire and J.~G.~Wacker,
%``The minimal moose for a little Higgs,''
arXiv:hep-ph/0206020.
%%CITATION = HEP-PH 0206020;%%

%\cite{Gregoire:2002ra}
\bibitem{Gregoire:2002ra}
T.~Gregoire and J.~G.~Wacker,
%``Mooses, topology and Higgs,''
arXiv:hep-ph/0206023.
%%CITATION = HEP-PH 0206023;%%

%\cite{Arkani-Hamed:2002qy}
\bibitem{Arkani-Hamed:2002qy}
N.~Arkani-Hamed, A.~G.~Cohen, E.~Katz and A.~E.~Nelson,
%``The littlest Higgs,''
JHEP {\bf 0207}, 034 (2002)
%%[arXiv:hep-ph/0206021].
%%CITATION = HEP-PH 0206021;%%

%\cite{Low:2002ws}
\bibitem{Low:2002ws}
I.~Low, W.~Skiba and D.~Smith,
%``Little Higgses from an antisymmetric condensate,''
Phys.\ Rev.\ D {\bf 66}, 072001 (2002)
[arXiv:hep-ph/0207243].
%%CITATION = HEP-PH 0207243;%%

%\cite{Kaplan:2003uc}
\bibitem{Kaplan:2003uc}
D.~E.~Kaplan and M.~Schmaltz,
%``The little Higgs from a simple group,''
arXiv:hep-ph/0302049.
%%CITATION = HEP-PH 0302049;%%

%\cite{Wacker:2002ar}
\bibitem{Wacker:2002ar}
J.~G.~Wacker,
%``Little Higgs models: New approaches to the hierarchy problem,''
arXiv:hep-ph/0208235.
%%CITATION = HEP-PH 0208235;%%

%\cite{Schmaltz:2002wx}
\bibitem{Schmaltz:2002wx}
M.~Schmaltz,
%``Physics beyond the standard model (Theory): Introducing the little  Higgs,''
arXiv:hep-ph/0210415.
%%CITATION = HEP-PH 0210415;%%

%\cite{Nelson:2003aj}
\bibitem{Nelson:2003aj}
A.~E.~Nelson,
%``Dynamical electroweak superconductivity from a composite little Higgs,''
arXiv:hep-ph/0304036.
%%CITATION = HEP-PH 0304036;%%

%\cite{Chivukula:2002ww}
\bibitem{Chivukula:2002ww}
R.~S.~Chivukula, N.~Evans and E.~H.~Simmons,
%``Flavor physics and fine-tuning in theory space,''
arXiv:hep-ph/0204193.
%%CITATION = HEP-PH 0204193;%%

%\cite{Hewett:2002px}
\bibitem{Hewett:2002px}
J.~L.~Hewett, F.~J.~Petriello and T.~G.~Rizzo,
%``Constraining the littlest Higgs,''
arXiv:hep-ph/0211218.
%%CITATION = HEP-PH 0211218;%%

%\cite{Csaki:2002qg}
\bibitem{Csaki:2002qg}
C.~Csaki, J.~Hubisz, G.~D.~Kribs, P.~Meade and J.~Terning,
%``Big corrections from a little Higgs,''
arXiv:hep-ph/0211124.
%%CITATION = HEP-PH 0211124;%%

%\cite{Chang:2003un}
\bibitem{Chang:2003un}
S.~Chang and J.~G.~Wacker,
%``Little Higgs and custodial SU(2),''
arXiv:hep-ph/0303001.
%%CITATION = HEP-PH 0303001;%%

%\cite{Csaki:2003si}
\bibitem{Csaki:2003si}
C.~Csaki, J.~Hubisz, G.~D.~Kribs, P.~Meade and J.~Terning,
%``Variations of Little Higgs Models and their Electroweak Constraints,''
arXiv:hep-ph/0303236.
%%CITATION = HEP-PH 0303236;%%

%\cite{Kribs:2003yu}
\bibitem{Kribs:2003yu}
G.~D.~Kribs,
%``Electroweak precision tests of little Higgs theories,''
arXiv:hep-ph/0305157.
%%CITATION = HEP-PH 0305157;%%

%\cite{Burdman:2002ns}
\bibitem{Burdman:2002ns}
G.~Burdman, M.~Perelstein and A.~Pierce,
%``Collider tests of the little Higgs model,''
arXiv:hep-ph/0212228.
%%CITATION = HEP-PH 0212228;%%

%\cite{Han:2003wu}
\bibitem{Han:2003wu}
T.~Han, H.~E.~Logan, B.~McElrath and L.~T.~Wang,
%``Phenomenology of the little Higgs model,''
arXiv:hep-ph/0301040.
%%CITATION = HEP-PH 0301040;%%

%\cite{Grinstein:1991cd}
\bibitem{Grinstein:1991cd}
B.~Grinstein and M.~B.~Wise,
%``Operator analysis for precision electroweak physics,''
Phys.\ Lett.\ B {\bf 265}, 326 (1991).
%%CITATION = PHLTA,B265,326;%%

%\cite{Burgess:1993vc}
\bibitem{Burgess:1993vc}
C.~P.~Burgess, S.~Godfrey, H.~Konig, D.~London and I.~Maksymyk,
%``Model independent global constraints on new physics,''
Phys.\ Rev.\ D {\bf 49}, 6115 (1994)
[arXiv:hep-ph/9312291].
%%CITATION = HEP-PH 9312291;%%

%\cite{Bamert:1996px}
\bibitem{Bamert:1996px}
P.~Bamert, C.~P.~Burgess, J.~M.~Cline, D.~London and E.~Nardi,
%``R_b and New Physics: A Comprehensive Analysis,''
Phys.\ Rev.\ D {\bf 54}, 4275 (1996)
[arXiv:hep-ph/9602438].
%%CITATION = HEP-PH 9602438;%%

%\cite{Chanowitz:2001bv}
\bibitem{Chanowitz:2001bv}
M.~S.~Chanowitz,
%``The Z $\to$ anti-b b decay asymmetry: Lose-lose for the standard model,''
Phys.\ Rev.\ Lett.\  {\bf 87}, 231802 (2001)
[arXiv:hep-ph/0104024].
%%CITATION = HEP-PH 0104024;%%

%\cite{Chanowitz:2002cd}
\bibitem{Chanowitz:2002cd}
M.~S.~Chanowitz,
%``Electroweak data and the Higgs boson mass: A case for new physics,''
Phys.\ Rev.\ D {\bf 66}, 073002 (2002)
[arXiv:hep-ph/0207123].
%%CITATION = HEP-PH 0207123;%%

%\cite{Barbieri:1999tm}
\bibitem{Barbieri:1999tm}
R.~Barbieri and A.~Strumia,
%``What is the limit on the Higgs mass?,''
Phys.\ Lett.\ B {\bf 462}, 144 (1999)
[arXiv:hep-ph/9905281].
%%CITATION = HEP-PH 9905281;%%

%\cite{Peskin:2001rw}
\bibitem{Peskin:2001rw}
M.~E.~Peskin and J.~D.~Wells,
%``How can a heavy Higgs boson be consistent with the precision  electroweak measurements?,''
Phys.\ Rev.\ D {\bf 64}, 093003 (2001)
[arXiv:hep-ph/0101342].
%%CITATION = HEP-PH 0101342;%%

\end{thebibliography}
\end{document}